\documentclass{ws-rv9x6}  

\newcommand{\ba}{\begin{eqnarray}}
\newcommand{\ea}{\end{eqnarray}}
\begin{document}

\chapter*{Electromagnetic radiation from relativistic nuclear
collisions} 

\author{Charles Gale$^{\dag}$ and Kevin. L. Haglin$^\S$}


\address{$^\dag$Department of Physics, McGill University\\
         3600 University St., Montr\'eal, QC, H3A 2T8, Canada\\}

\vspace*{-0.6cm}

\address{$^{\S}$Department of Physics, Astronomy and Engineering Science\\
St. Cloud State University, St. Cloud, MN 56301, USA\\}

\date{\today}

\begin{abstract}
We review some of the results obtained in the study of the production of 
electromagnetic radiation in relativistic
nuclear collisions. We concentrate on the emission of real photons and dileptons
from the hot and dense strongly interacting phases of the reaction.  We
consider the contributions from the partonic sector, as well as those from the
nonperturbative hadronic sector. We examine the current data, some of the
predictions for future measurements, and comment on
what has been learnt so far. 

\end{abstract}

\newpage
\tableofcontents
\newpage

\section{Introduction}

The study of matter under extreme conditions constitutes a rich field of 
intellectual pursuit and is a vibrant research area of physics.  
It is popularised to nonspecialists by indicating that such studies
reveal physics that governed the early Universe (microseconds after the 
Big Bang) and also continue today to govern physics of compact astrophysical 
objects (neutron stars and black holes).  But it is indeed more than
that.  Practitioners concern themselves with a variety of very specific and 
technically challenging questions.  When ordinary matter is heated to roughly
a trillion degrees Kelvin, how does it respond?  And what are the hallmark signatures 
of this response?  What should one look for?  When relatively cold matter is 
compressed to densities many times greater
than that of normal nuclei, does it resemble something other than ordinary
protons and neutrons?   Quantum Chromodynamics (QCD) predicts under
these extreme conditions, very far from the ground state, that matter
will essentially change its properties to resemble a plasma of 
quarks and gluons (QGP)\cite{lattice,kanaya}.  What is the QCD phase diagram?

\begin{figure}
\begin{center}
\includegraphics[width=6cm]{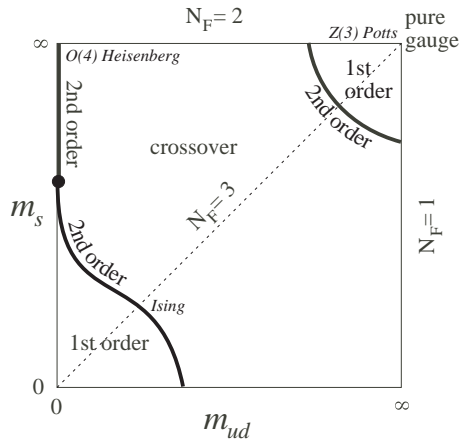}
\caption{The nature of the QCD phase transition as a function of quark masses,
along with theoretical expectations from effective models. From
[\protect\refcite{kanaya}].}
\label{Columbia}
\end{center}
\end{figure}

Indeed, the understanding of QCD under extreme conditions of high 
temperature or large baryon 
density has progressed considerably in recent years. Confinement, in the pure
glue version of QCD, is a property that can be associated with a definite
symmetry whose status is probed by the value of the Polyakov loop $\langle L
\rangle$. This symmetry is valid at low temperature, but broken at high
temperatures. In the limit of massless quarks, QCD is chirally symmetric, and
that symmetry is valid at high temperatures and spontaneously broken at low
temperatures. The order parameter there is the chiral condensate: $\langle
\bar{\psi} \psi \rangle$. As a function of temperature, those order parameters
are best studied on the lattice; although it is fair to say that the lattice
has just started to venture into the finite baryon domain with any degree of
quantitative assurance\cite{fokar}. An idea on the status of finite-temperature lattice QCD
can be had by consulting Ref.~[\refcite{lattice,kanaya}]. The order of the transition
actually depends on the details of the parameters of the theory, as shown in the
``Columbia plot'' (Fig.~\ref{Columbia}), but even in the
absence of a robust prediction at finite temperature and baryon density, the 
lattice does provide tantalising clues of an eventually observable behaviour of
the many-body nature of QCD. For example, it offers some support, together with
effective models, to the discussions of a genuine tricritical point in the QCD
phase diagram\cite{tricrit}. 

The experimental realisation of hot and dense strongly interacting matter in
terrestrial accelerators has been accomplished through the study of 
relativistic nuclear collisions.  Producing quark gluon plasma
(QGP) is one of the premier goals of the Relativistic Heavy 
Ion Collider (RHIC) and its broader experimental program
and represents already enormous challenge; identifying the
plasma and also studying its unique
properties is yet something else, and has proven to be extremely challenging
from both the theoretical and experimental points of view. 
Signatures of the QGP have been proposed:
strangeness enhancement, suppression of the $J/\psi$ signal,
effects of multiple collisions on the observed particle spectra, and 
electromagnetic radiation are but a few.  Unfortunately, while most of the 
proposed signatures are plausible and evidently do occur at some level, it has
been difficult to refine the mostly heuristic arguments into really precise
predictions. It is now clear that certainty will be attained in this field
through the simultaneous analysis of complementary observables. This being said,
probes that do not interact strongly have the definite advantage of suffering
little or no final state interaction: they open a privileged window to the hot
and dense phases of the reaction. The price to pay is a small production
rate.

We will discuss in this  
article the status of theory and the current picture relative
to the experimental data, surrounding electromagnetic radiation as probes
of strongly-interacting many-body dynamics.
This work is organised as follows: a brief review of the formalism germane 
to the emission of real and virtual photons from hot and dense systems is 
followed by a
discussion of the radiation from the hadronic sector, then from the partonic
sector. Some of predictions for current and future measurements are outlined,
followed by a conclusion. The main thrust here pertains to models amenable to
conventional experimental measurements: the high density low temperature phase
of QCD\cite{raj} will not be discussed, even though its electromagnetic emissivity
has been investigated\cite{prash}. A goal here is to provide the reader
with a snapshot of this rapidly evolving field by discussing some of the recent
theoretical and experimental developments. In doing so, it is unfortunately
impossible to do justice to the wholeness of the body of work in this exciting
area: we shall be brief on some topics and refer instead to the literature and
in particular to previous reviews.

\section{Radiation from Hadronic Matter}
\label{impart}

The goal of this section of the text is to relate the spectrum of emitted
radiation to some of the intrinsic properties of the strongly interacting
matter. A thermalised medium is assumed, and the formalism below is developed in
the one-photon approximation. 

Quite generally, the electromagnetic radiation from strongly interacting
matter can be related to the imaginary part of the retarded in-medium 
photon self-energy, at finite temperature and 
density\cite{fein76,mcle85,ko89,weld90,galk91}. We sketch a  
derivation\cite{galk91} here, for the emission of real photons.  Consider a 
transition $i \to f \gamma$, {\it\/i.e.} from some initial hadronic 
state $i$ to some final
hadronic state $f$, plus a photon of momentum $k^\mu = (\omega, {\bf k})$
and polarisation $\epsilon^\mu$.  The transition rate is
\ba
R_{f i} = \frac{| S_{f i}|^2}{\tau V}\/,
\ea
where $\tau$ is the observation time, $V$ is the volume of the system, 
and the $S$-matrix element is 
\ba
S_{f i} = \langle f | \int d^4 x \, J_\mu (x) A^\mu (x) \, | i \rangle\,.
\ea
$J_\mu (x)$ is an electromagnetic current operator, and 
\ba
A^\mu (x) = \frac{\epsilon^\mu}{\sqrt{2 \omega V}} (e^{i k \cdot x} + 
e^{ - i k \cdot x})\,.
\ea
Using translation invariance, summing over polarisations, and using the 
integral representation of the delta function, one can write
\ba
R_{f i} &=& - \frac{g^{\mu \nu}}{2 \omega V} (2 \pi)^4 \left[ \delta^4 (p_i 
+ k - p_f) + \delta^4 (p_i - k - p_f) \right]
\nonumber\\
&\ &\quad\times \langle f | J_\mu (0) | i \rangle \, \langle 
i | J_\nu (0) | f \rangle\, ,
\ea
where the parts having to do with emission and absorption are evident. The 
net thermal rate is obtained by summing the above over final states and 
taking a thermal average of the initial configurations. 

It is useful at this point to define some finite-temperature 
current-current correlators\cite{frad65}:
\ba
f^>_{\mu \nu} (k) = \int d^4 x\,e ^{i k \cdot x} \sum_i \langle 
i | A_\mu (x) A_\nu (0)
| i \rangle\, e^{-\beta E_i} / Z\,,\nonumber \\
f^<_{\mu \nu} (k) = \int d^4 x\,e ^{i k \cdot x} \sum_i \langle 
i | A_\mu (0) A_\nu (x)
| i \rangle\, e^{-\beta E_i} / Z\,,\nonumber \\
f^{\rm R}_{\mu \nu} (k) = \int d^4 x\,e ^{i k \cdot x} \sum_i \langle 
i |\, [ A_\mu (x), 
A_\nu (0) ]\, | i \rangle\, e^{-\beta E_i} / Z\,.
\ea
The last correlation function is a retarded correlation function. The 
above all involve the
electromagnetic current operator in the Heisenberg picture  and are
written in the grand canonical ensemble, where $Z$ is the partition 
function. Assuming translational invariance, 
the first two can be rewritten together as
\ba
f^{\stackrel{\textstyle >}{<}}_{\mu \nu} (p^0, \vec{p}\,) = \sum_{i, f} 
(2 \pi)^4 \delta^4 (p_f - p_i \pm p) \langle i | A_\mu (0)
| f \rangle \, \langle f | A_\nu (0) | i \rangle e^{-\beta E_i}/Z\,.
\ea
Clearly, $f^>$ is involved with absorption of radiation, whereas $f^<$ 
deals with emission.  Only, the latter case is treated here.  Further defining 
a spectral density 
$\rho_{\mu \nu} = f^>_{\mu \nu} -
f^<_{\mu \nu}$, one may first show that 
\ba
f^<_{\mu \nu} = \frac{\rho_{\mu \nu}}{e^{\beta \omega} - 1}\,,
\ea
and also that 
\ba
f^{\rm R}_{\mu \nu} = i \int \frac{d \omega}{2 \pi}\, 
\frac{\rho_{\mu \nu}(\omega, \vec{k}\,)}
{k^0 - \omega + i \epsilon}\,.
\ea
With the above elements, one can finally write $\rho_{\mu \nu} = 
2 \,{\rm Im} f^{\rm R}_{\mu \nu}$.
The emission probability is related to the imaginary part of the 
finite-temperature retarded 
current-current correlation function. In  the one-photon  
approximation ({\it i.e.} to lowest order
in $e^2$), the time-ordered current correlator is the one-particle 
irreducible photon self-energy, $\Pi_{\mu \nu}$\cite{ps95}. Putting 
all of this together, the differential rate for emitting real photons is
\ba
\omega \frac{d^{\/3}R}{d^{\/3}k} = - \frac{g^{\mu \nu}}{( 2 \pi)^3} \, 
{\rm Im}\Pi^{\rm R}_{\mu \nu} (k) \frac{1}{e^{\beta \omega} - 1}\,.
\ea
The proof is easily generalised to the case of lepton pair emission 
(the lepton mass has been set to zero):
\ba
E_+ E_- \frac{d^6 R}{d^3 p_+ d^3 p_-} &=& \frac{2 e^2}{(2 \pi)^6} 
\frac{1}{k^4} \left[ p^\mu_+
p^\nu_- + p^\nu_+ p^\mu_- - g^{\mu \nu} p_+ \cdot p_- \right]\, 
{\rm Im} \Pi^{\rm R}_{\mu \nu} (k)
\nonumber\\
 & \ &\quad\times
\frac{1}{e^{\beta \omega} - 1}\,.
\ea

Thus, the electromagnetic signal emitted during a nuclear reactions can be
related to a quantity that is linked to properties of the medium itself. As it
shall be seen, the retarded self-energy (or alternatively the current-current
correlator) will be modified in a strongly interacting environment. Also, the
current-current correlator is calculable only perturbatively, unless a specific
model is available: this again testifies to the importance and the value of
electromagnetic measurement in nuclear collisions. They indeed open a window to
the hot and dense phases of the reaction, and those regions can't be probed directly
by other means. 

\subsection{The Low Dilepton Invariant Mass Sector}

\subsubsection{A Baseline Calculation}

\begin{figure}
\begin{center}
\includegraphics[width=5cm]{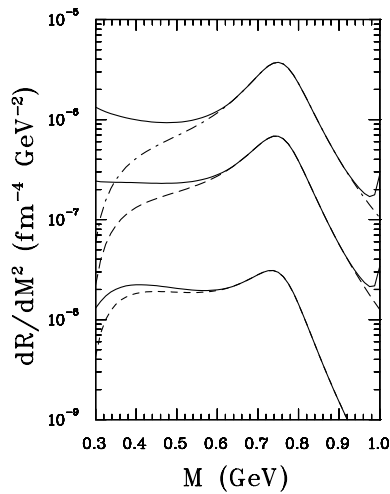}
\caption{The rates for dielectron emission from an incoherent sum of 
meson reactions\protect\cite{gali94}.  The three sets of full curves are
the net rates at a temperature of 100, 150, and 200 MeV, from bottom to 
top. The dashed curves represent the $\pi^+ \pi^- \to e^+ e^-$ contribution
only.}
\label{galirates}
\end{center}
\end{figure}

As a calculation to set the scale of the physical processes under 
consideration, it is useful to
consider first the following question: what is the magnitude of the 
radiation emitted by a hot gas
of mesons? Specialising to the lepton pair sector\cite{gali94}, this 
problem is briefly summarised here. Using relativistic kinetic theory, 
the lepton pair emission rates can be
calculated with the help of effective interaction Lagrangians. The 
parameters of those effective
Lagrangians are fitted to radiative decays measurements using
vector meson dominance (VMD). 
Specifically, the
``calibration'' reactions are: $\rho \to \pi \gamma$, $K*^\pm \to K^\pm 
\gamma$, $K*^0(\bar{K*}^0)
\to K^0 (\bar{K*}^0)\, \gamma$, $\omega \to \pi^0 \gamma$, $\rho^0 \to 
\eta \gamma$, $\eta' \to
\rho^0 \gamma$, $\eta' \to \omega \gamma$, $\phi \to \eta \gamma$, $\phi 
\to \eta' \gamma$, and
$\phi \to \pi^0 \gamma$. Note that the usage of relativistic kinetic 
theory here is tantamount to evaluating the finite temperature photon 
self-energy at the one-loop level\cite{galk91}. The rate for $a b 
\to e^+ e^-$ is given by
\ba
R_{a b \to e^+ e^-} &=& {\cal N} \int \frac{d^3 p_a}{2 E_a 
(2 \pi)^3} \frac{d^3 p_b}{2 E_b (2 \pi)^3}
\frac{d^3 p_+}{2 E_+ (2 \pi)^3} \frac{d^3 p_-}{2 E_- (2 \pi)^3} f_a f_b 
\nonumber\\
& \ & \times |\bar{\cal{\/M\;}}|^2 (2 \pi)^4
\delta^4 (p_a + p_b - p_+ - p_-)\ ,
\ea
where the $f$'s are appropriate distribution functions, and ${\cal N}$ 
is an overall degeneracy factor dependent upon the specific channel. 
The $P P \to e^+ e^-$, $P V \to e^+ e^-$, and $V V \to
e^+ e^-$ reactions can be obtained from the radiative decay ones through 
crossing symmetry. The sum
of them is shown in Fig.~\ref{galirates}. Also shown is the contribution 
from the $\pi^+ \pi^- \to
e^+ e^-$ reaction. This channel has often been considered as the sole 
source of hadronic dileptons
in early calculations, owing mainly to multiplicity arguments. Even in this 
incoherent sum approach, one can see that this assumption is badly violated 
in the low mass region.

\subsubsection{Spectral Density Calculations}

The full power of the formalism derived in section \ref{impart} reveals 
itself when it is made clear that the electromagnetic radiation rate is
related to the in-medium vector meson spectral density: this quantity is 
of course not measurable directly. The spectral density is related
to the imaginary part of the full propagator, by definition. It appears when the 
electromagnetic current operator in the retarded self-energy is expressed as 
the field operator, through VMD\cite{sak,weise,ocon}. We illustrate the spectral 
density approach here, by describing a calculation again done in the
baryonless regime\cite{raga99}. 

One may start with the model for the $\rho$-meson in free space
employed previously in Refs.~[\refcite{CS93,CRW,RCW}]. 
Based on the standard $\rho\,\pi\pi$ interaction vertex (isospin structure 
suppressed), 
\begin{equation}
{\cal\/L}_{\rho\pi\pi} = g_{\rho\pi\pi} \ \pi \ p^\mu \pi
\rho_\mu \ ,
\label{Lrhopipi}
\end{equation} 
($p^\mu$: pion momentum) the bare $\rho$-meson
of mass $m_\rho^{0}$ is renormalised through the two-pion loop
including a once-subtracted dispersion relation, 
giving rise to the vacuum self-energy
\begin{eqnarray}
\Sigma_{\rho\pi\pi}^0(M) & = & \bar{\Sigma}_{\rho\pi\pi}^0(M)
-\bar{\Sigma}_{\rho\pi\pi}^0(0) \ ,
 \nonumber\\
\bar{\Sigma}_{\rho\pi\pi}^0(M) & = & \int \frac{p^2 dp}{(2\pi)^2} \
v_{\rho\pi\pi}(p)^2 \ G_{\pi\pi}^0(M,p)  \  ,
\label{sigrho0}
\end{eqnarray}
with the vacuum two-pion propagator
\begin{equation}
G_{\pi\pi}^0(M,p)=\frac{1}{\omega_\pi(p)} \
\frac{1}{M^2-(2\omega_\pi(p))^2+i\eta}
 \ ; \quad \omega_\pi(p)=\sqrt{m_\pi^2+p^2}\,, \
\label{Gpipi}
\end{equation}
and vertex functions 
\begin{equation}
v_{\rho\pi\pi}(p) 
= \sqrt{\frac{2}{3}} \ g_{\rho\pi\pi} \ 2p \ F_{\rho\pi\pi}(p)\,, 
\label{vrhopipi}
\end{equation}
involving a hadronic (dipole) form factor $F_{\rho\pi\pi}$~\cite{RCW}.
Resumming the two-pion loops in a Dyson equation gives the free $\rho$
propagator
\begin{equation}
D_\rho^0(M)=[M^2-(m_\rho^{0})^2-\Sigma_{\rho\pi\pi}^0(M)]^{-1} \ ,
\end{equation}
which agrees well with the measured $p$\,-wave $\pi\pi$ phase shifts 
and the pion electromagnetic form factor obtained within VMD. 
\begin{figure}[h]
\begin{center}
\includegraphics[width=6cm]{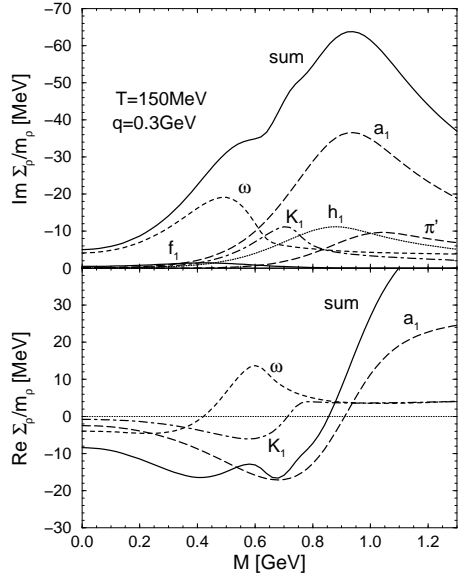}
\caption{The real and imaginary parts of the polarisation-averaged
$\rho$ self-energy (lower and upper panel, respectively). The
different channels are labelled explicitly and are explained in the
text\protect\cite{raga99}.
Note that the $\pi\pi$ channel is absent for the sake of viewing clarity.}
\label{rhoself}
\end{center}
\end{figure}
\begin{figure}[h]
\begin{center}
\includegraphics[width=6cm,angle=-90]{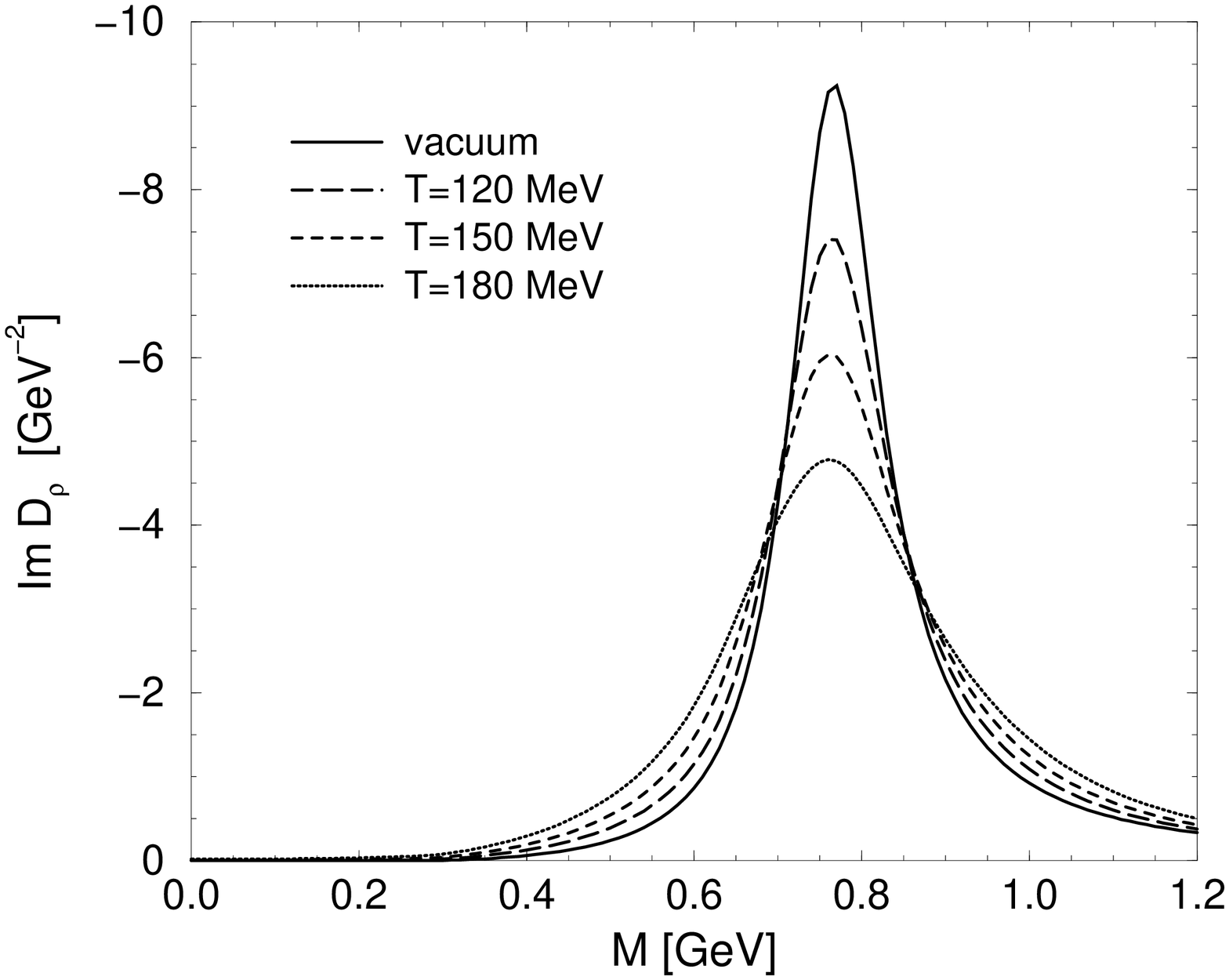}
\caption{Imaginary part of the $\rho$ propagator in the vacuum (solid
curve) and in a thermal gas including the full in-medium self-energies
for fixed three-momentum $q$ = 0.3 GeV at temperatures $T$ =
120 MeV (long-dashed curve), $T$ = 150 MeV (dashed curve), and
$T$ = 180 MeV (dotted curve).  The figure comes from 
Ref.~[\protect\refcite{raga99}].}
\label{spectral}
\end{center}
\end{figure}

To calculate medium corrections to the $\rho$ self-energy in a hot meson
gas, one can assume that the interactions are dominated
by $s$-channel resonance formation. It is then possible to use the formal
relationship that relates the self-energy to the forward scattering amplitude,
integrated over phase space at finite temperature. A field-theoretic derivation
of this connection can be found in [\refcite{jeon}]. 
At moderate temperatures relevant
for the hadronic gas phase, the light pseudoscalar Goldstone bosons
$P=\pi,K$ are the most abundant species.
The various resonances in $\rho P$ collisions can be grouped into
two major categories, namely vector mesons $V$ and
axial-vector mesons $A$. The effective Lagrangians that regulate the
interactions among all those species have their parameters chosen such that
the measured hadronic phenomenology are reproduced\cite{raga99}. This statement
also holds true for hadronic form factors. In order to calculate the $\rho$
spectral density at moderate temperatures, one includes the hadronic fields
appearing in Table~\ref{tableraga}. 
\begin{table}
\tbl{\label{tableraga}Mesonic resonances $R$ with masses $m_R\le 1300$~MeV
and substantial branching ratios
into final states involving direct $\rho$'s (hadronic)
or $\rho$\,-like photons (radiative). Taken 
from Ref.~[\protect\refcite{raga99}].}
{\begin{tabular}{c|ccccc}
 $R$ & $I^GJ^P$ & $\Gamma_{tot}$ [MeV] & $\rho h$ Decay &
$\Gamma^0_{\rho h}$ [MeV] & $\Gamma^0_{\gamma h}$ [MeV] \\
\hline
$\qquad\omega(782)\qquad$ & $0^-1^-$ & 8.43 & $\rho\pi$ & $\sim 5$ & 0.72 \\
$h_1(1170)$   & $0^-1^+$ & $\sim 360$  & $\rho\pi$ & seen  &   ?  \\
$a_1(1260)$   & $1^-1^+$ & $\sim 400$  & $\rho\pi$ & dominant & 0.64 \\
$K_1(1270)$   & $\frac{1}{2}1^+$ & $\sim 90$ & $\rho K$  & $\sim 60$ &   ?  \\
$f_1(1285)$   & $0^+1^+$        & 25 & $\rho\rho$ & $\le$8   & 1.65  \\
$\pi'(1300)$  & $1^-0^-$        & $\sim 400$ & $\rho\pi$ & seen   & ?  \\
\end{tabular}}
\end{table}
The interaction vertices being completely determined, one may first calculate the
in-medium $\rho$ self-energy, then the complete propagator in terms of its 
longitudinal and transverse parts\cite{galk91}. The real and imaginary parts of
the in-medium $\rho$ self-energy are shown in Fig.~\ref{rhoself}. Each curve is
labelled according to that species which interacts with the $\rho$. It is
instructive to observe that the imaginary parts all add, while there is a
significant amount of cancellation of the real parts. The first effect creates
a sizeable width for the in-medium $\rho$, while the second determines the
in-medium mass, which appears to be only slightly modified. Those different
aspects are again seen in the representation of the imaginary part of the
in-medium $\rho$ propagator, shown in Fig.~\ref{spectral}. 

The calculations of the in-medium vector meson spectral densities clearly show
the richness of the many-body problem under scrutiny. 
The power of the techniques
described above becomes evident when they are combined with dynamical models and
confronted with experimental data. This story is well chronicled in 
[\refcite{cbrat99,rw00}], and in references therein. See also [\refcite{hbek02}]. The 
current situation can be
summarised by writing that the low dilepton invariant mass
data\cite{ceres,masera} can be understood in terms of in-medium modifications
of vector spectral densities. These data can not empirically exclude, 
however, other interpretations\cite{brown_rho,cbrat99,galeqm01}.
\begin{figure}
\begin{center}
\includegraphics[width=6cm]{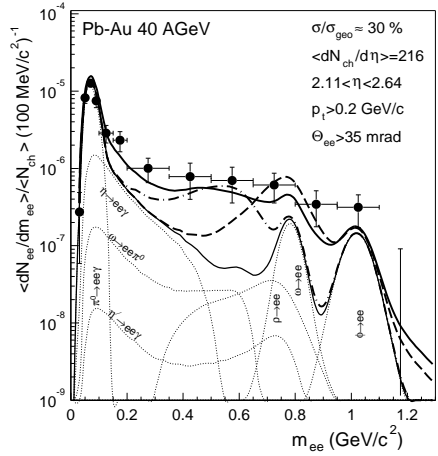}
\caption{Dilepton spectrum from Pb-Au collisions at 40 AGeV/c. See the text for
the meaning of the different curves. Data are from [\protect\refcite{JW02}].}
\label{ceres_low}
\end{center}
\end{figure}
This
unfortunate situation still prevails at a lower energy\cite{JW02}, as shown in
Fig.~\ref{ceres_low}. The sources there are: free hadron decays without $\rho$
decay (thin solid line), calculation with a vacuum $\rho$ spectral density
(thick dashed line), dropping in-medium $\rho$ mass\cite{brown_rho} (dash-dotted
line), and with a medium-modified $\rho$ spectral density\cite{rw00} (thick solid
line). 
Note that 
but the persistence of the dilepton excess at lower energies does support a baryon
density-driven effect. Other suggestions to resolve the different models require
high statistics\cite{cbrat99}: the final analysis of the CERES 2000 data with
the TPC is eagerly anticipated. 

\subsection{The Dilepton Intermediate Invariant Mass Sector}
\label{interm_sect}

\begin{figure}
\begin{center}
\includegraphics[width=8cm]{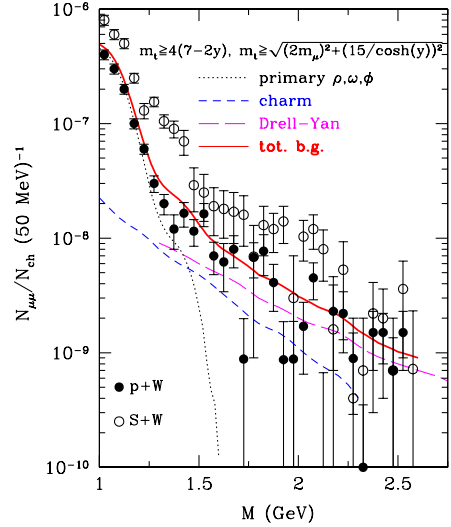}
\caption{Comparison of lepton-pair yield divided by the multiplicity of charged 
particles, 
in p + W and S + W collisions at 200 A GeV/c. The data are from [\protect\refcite{masera}].}
\label{heliosdata}
\end{center}
\end{figure}

Intermediate mass lepton pairs have traditionally been a focus of interest,
as their spectrum has been suggested early on as a signature of the 
quark-gluon plasma\cite{shu78}.  In relativistic nuclear collisions, 
measurements have been carried out at SPS energies by the HELIOS-3 and 
the NA38/NA50 collaborations in the lepton pair invariant mass 
range $m_{\phi} < M < m_{J/\psi}$. Both experimental collaborations have 
observed significant enhancement of dilepton
yield in this region for central S + W and S + U collisions as compared to
those in proton-induced reactions (normalised to the charged-particle
multiplicity)\cite{helios3,na38}. Chronologically, HELIOS-3  reported on the
intermediate-mass enhancement first. This experiment was designed to study
virtual photons in the dimuon sector at low transverse mass. In this way,
dimuon production was studied from threshold up the $J/\psi$ mass over a
wide range in $p_{\rm\/T}$. A good summary of this experimental situation is
shown in Fig.~\ref{heliosdata}. 
Several explanations of the intermediate mass dimuon enhancement have been put
forward. Those include additional production of 
$c \bar{c}$ pairs\cite{ccbar}, secondary Drell-Yan emission\cite{dy2}, and
charmed meson rescattering\cite{lw98}. Note that in principle, all of those 
effects can coexist.  However, such a global modelling has not been done, and 
we thus discuss them separately. It is fair to say that the first of those 
mentioned above
is still admitted by current experimental data\cite{capelli}, even though 
some calculations are slightly less optimistic\cite{lmw}. The role of the last 
ingredient on the list has not been found to be large enough to account for 
the experimentally observed excess\cite{bordalo,capelli}. 

Another class of approaches consists of quantitative evaluation of
thermal dilepton sources. Those may be from the hadronic, confined sector of
QCD, and/or from the quark gluon plasma itself. One such model is described below.
Recall that thermal hadronic sources have been shown to be crucial in the low
mass sector. It is therefore legitimate to ask how high in invariant mass is
the extent of the virtual photon radiation from those sources?  Those 
concerns are carried to their logical conclusion in what follows. 

In the intermediate invariant mass region, relativistic theory estimates
indicate that the following microscopic channels are relevant: $\pi \pi \to
l^+ l^-$, $\pi \rho \to l^+ l^-$, $\pi \omega \to l^+ l^-$, $\pi a_1 \to l^+
l^-$, $K \bar{K} \to l^+ l^-$, and $K \bar{K}^* + {\rm c.c.} 
\to l^+ l^-$ [\refcite{gali}]. Apart from sheer coupling constant values, the
importance of those contributions stems simply from considering energy 
scales involved and from phase space arguments.  This combination of  
coupling constants and phase space is effective in maximising a
particular contribution from the $\pi a_1$ channel, for
example\cite{pia1}. 

\begin{figure}
\begin{center}
\includegraphics[width=8cm]{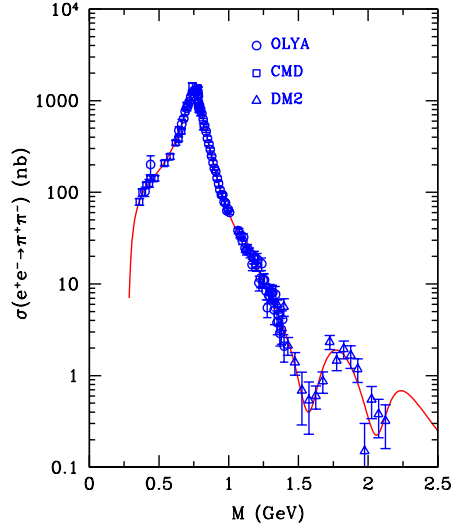}
\caption{The cross section for $e^+ e^- \to \pi^+ \pi^-$. The solid curve is
based on the model of Ref.~[\protect\refcite{piff}]. The experimental data 
are from the OLYA collaboration\protect\cite{olya}, the CMD 
collaboration\protect\cite{olya}, and the DM2 collaboration 
\protect\cite{dm2}.} 
\label{piform}
\end{center}
\end{figure}

\begin{figure}
\begin{center}
\includegraphics[width=8cm]{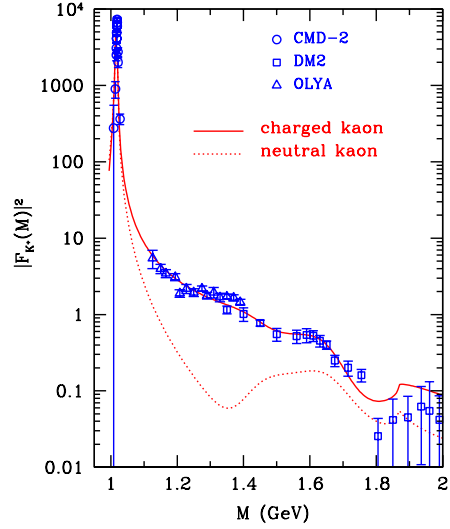}
\caption{The kaon electromagnetic form factor. The solid and dotted 
curves\protect\cite{piff} are
for the charged and neutral kaons, respectively. The symbols are for the
charged kaon data from the CMD-2 collaboration\protect\cite{cmd2}, the DM2
collaboration\protect\cite{dm2_2}, and the 
OLYA collaboration\protect\cite{olya_2}.}
\label{kff}
\end{center}
\end{figure}

\begin{figure}
\begin{center}
\includegraphics[width=8cm]{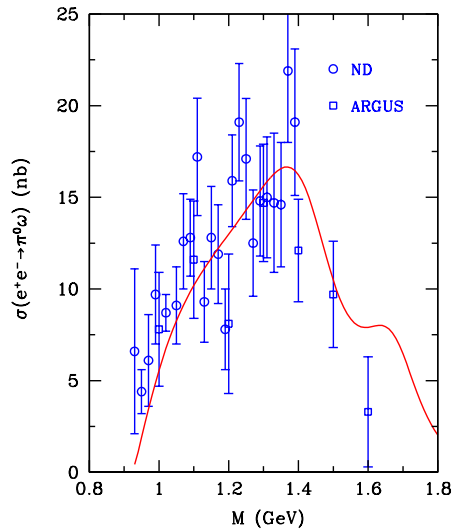}
\caption{The cross section for $e^+ e^- \to \pi^0 \omega$. The solid curve is
described in the text. The experimental data are from the ND\protect\cite{nd}
and ARGUS\protect\cite{argus} collaborations. }
\label{piom}
\end{center}
\end{figure}

Calculations of dilepton-emitting processes in the intermediate invariant mass
region follow similar steps to those in the low mass sector. Effective
Lagrangians are used, together with VMD, and the coupling constants and
possible form factors are fitted to measured 
strong decays and
electromagnetic radiative decays. Only, in the intermediate mass domain an
extrapolation is required. The strong decay widths set a scale that is
typically an order of magnitude below the mass region of interest: 1 GeV $<$
$M$ $<$ 3 GeV. The radiative decays are even smaller, owing to the size of
$\alpha_{\rm em}$ [\refcite{pdg}]. The required extrapolation is then 
vulnerable to off-shell effects. Put another way, there is a risk of 
uncontrolled growth of form factors since the application region is far 
removed from the region where the empirical fitting was realized. Indeed, 
different Lagrangians known to agree in the low mass sector generating
dileptons were found to differ significantly in their predictions of 
emission rates 
for intermediate mass lepton pairs\cite{gaogale}. Fortunately, there 
exists a wealth of data for $e^+ e^- \to {\rm hadrons}$, exactly in the 
invariant mass window relevant for this application. Those data 
can thus be used to extract an effective form factor for the inverse 
reactions. Alternatively, they may also be used to extract spectral 
densities: this point will be discussed later. As an example, consider 
$e^+ e^- \to \pi^+ \pi^-$, which has been measured with high 
accuracy\cite{olya,dm2}, evidenced by the data shown in Fig.~\ref{piform}.
The cross section for this reaction can be written as
\begin{eqnarray}
\sigma ( e^+ e^- \to \pi^+ \pi^-) = \frac{8 \pi \alpha^2 k^3}{3 M^5} |F_\pi
(M)|^2\ ,
\end{eqnarray}
where $k$ is the three-momentum in the two-body rest frame, $M$ is the
lepton-pair invariant mass, and $F_{\pi}$ is the time-like pion electromagnetic
form factor. With these data, one can extract $F_\pi$, and then use it in
the calculation of the dilepton-producing reaction $\pi^+ \pi^- \to l^+ l^-$:
\begin{eqnarray}
\sigma ( \pi^+ \pi^- \to l^+ l^- ) = \frac{8 \pi \alpha^2 k}{3 M^3} |F_\pi ( M
)|^2 \left(1 - \frac{m_l^2}{M^2}\right) \left(1 + \frac{2 m_l^2}{M^2}\right)\
,
\end{eqnarray}
where $m_l$ is the lepton mass. A similar procedure can be followed for other
channels. Another example appears in Fig.~\ref{kff}, where the time-like
electromagnetic form factors for the kaon systems have been extracted from
electron-positron annihilation data.
Both processes introduced above are of the pseudoscalar-pseudoscalar type. For
the pseudoscalar-vector class, in the invariant mass region of interest,
$\pi \rho \to l^+ l^-$, $\bar{K} K^* + {\rm c.c.} \to l^+ l^-$, and
$\pi \omega \to l^+ l^-$ are included. The first two processes effectively 
involve three
pions, while the third one involves four pions. Note that in a transport
approach, a process involving three or more pions in the initial state can
only be described as a two-step process with an intermediate resonance. The
first two channels above have been studied in Ref.~[\refcite{haga}]. The
effective form factors one extracts are shown in [\refcite{gali}]. Details 
about the $\pi \omega$ channel are gotten from the study of four-pion 
final states. Using a Wess-Zumino {\it\/VVP\/} interaction Lagrangian, one 
finds
\begin{eqnarray}
\sigma ( \pi^0 \omega \to l^+ l^-) = \frac{4 \pi \alpha^2 k}{9 M} |F_{\pi
\omega} (M)|^2\ ,
\end{eqnarray}
in the limit of vanishing lepton mass. The form factor may be parametrised in
terms of three isovector $\rho$-like vector mesons, $\rho(770)$, $\rho(1450)$,
and $\rho(1700)$\cite{nd}:
\begin{eqnarray}
F_{\pi \omega} (M) = \sum_V g_{V \pi \omega} \frac{e^{i \phi_V} m_V^2}
{\left(m_V^2 - M^2\right) - i m_V \Gamma_V} \ .
\end{eqnarray}
This form is then used to fit the experimental data from the ND and ARGUS
collaborations. The result is shown in Fig.~\ref{piom}.

\begin{figure}
\begin{center}
\includegraphics[width=8cm]{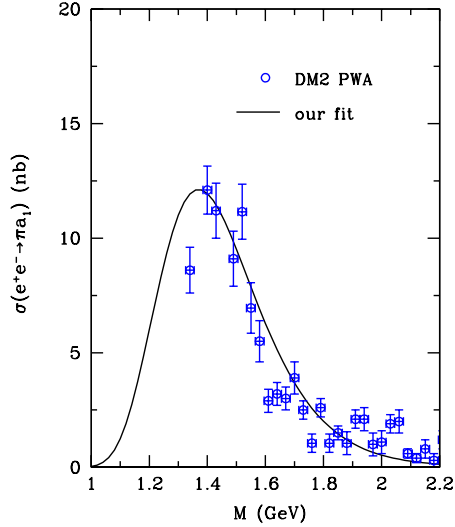}
\caption{The cross section for $e^+ e^- \to \pi a_1$. The open circles
are the experimental data from the DM2 collaboration using a 
partial wave analysis\protect\cite{pwa}. The solid curve is a fit to
the data.}
\label{pwafig}
\end{center}
\end{figure}

In the pseudoscalar axial-vector channel, we shall consider mainly $\pi a_1 \to
l^+ l^-$, which is in effect a four-pion process. Considering here again the
reaction where the lepton pair constitutes the initial state and the hadrons the
final, one can attempt an extraction of an effective form factor. Some previous
thermal rate calculations indicate that this specific channel is particularly
important in the intermediate mass region\cite{pia1}, even though it is
difficult to calculate reliably a specific signal using effective Lagrangians.
This fact owes mainly to off-shell effects\cite{gaogale}. One can pick a model
that yields adequate hadronic phenomenology on-shell, and then extrapolate 
to the intermediate mass sector with the help of experimental data. Using a 
chiral Lagrangian where the vector mesons are introduced as massive
Yang-Mills fields\cite{gomm} one may derive the following cross section 
\begin{eqnarray}
\sigma (\pi a_1 \to l \bar{l}\,) = \frac{\pi \alpha {\cal H\/}}{72 m_{a_1}^2 
g_\rho^2
M^5 k_\pi}\,|F_{\pi a_1}|^2 \left(1 - \frac{4 m_l^2}{M^2}\right)\, \left(1 +
\frac{2 m_l^2}{M^2}\right)\ ,
\end{eqnarray}
where ${\cal H}$ is a nontrivial function of coupling constants, masses and
momenta. $k_\pi$ is the magnitude of the pion momentum in the centre-of-mass. 
The issue of the electromagnetic form factor $|F_{\pi a_1}|^2$, can be
settled, at least in principle, by analysing $e^+ e^- \to \pi^+ \pi^- \pi^+
\pi^-$ and $e^+ e^- \to \pi^+ \pi^- \pi^0 \pi^0$ data. Although many
such analyses have been carried out, an unambiguous result is still elusive, 
as many other intermediate states may contribute to the same four-pion
final state. Several scenarios have been considered, and a discussion
appears in [\refcite{gali}].  What is probably a conservative estimate 
is highlighted here.  The DM2 collaboration has determined the cross
section $\sigma_{e^+ e^- \to \pi a_1}$ using a partial wave analysis 
(PWA)\cite{pwa}. One may extract an effective form factor from these 
data, see Fig.~\ref{pwafig}, and
then carry out an analysis for $\sigma_{{\pi a_1} \to l \bar{l}}$, using
detailed balance. 

\begin{figure}
\begin{center}
\includegraphics[width=8cm,angle=90]{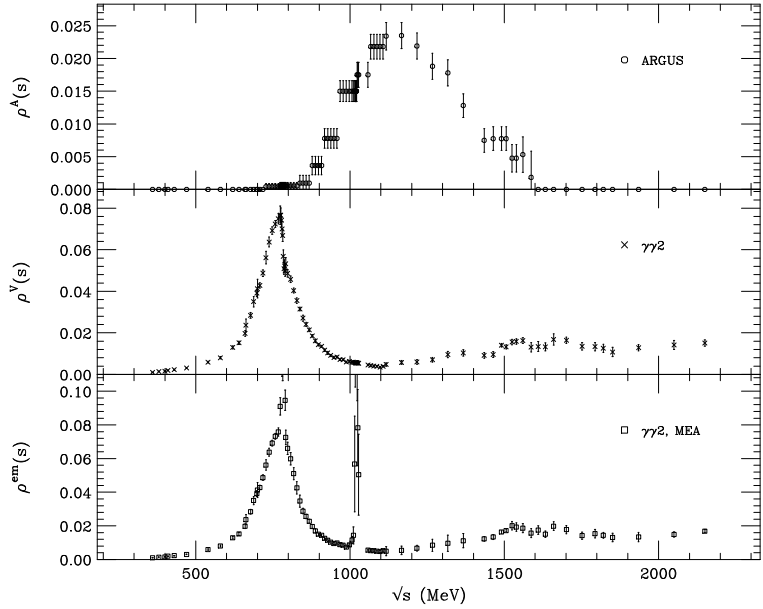}
\caption{The spectral functions $\rho^{em} (s)$, $\rho^V (s)$, and
$\rho^A (s)$, as compiled from $e^+ e^-$ annihilation and $\tau$ decay
data\protect\cite{huang}. }
\label{pwa}
\end{center}
\end{figure}

\begin{figure}
\begin{center}
\includegraphics[width=8cm,angle=0]{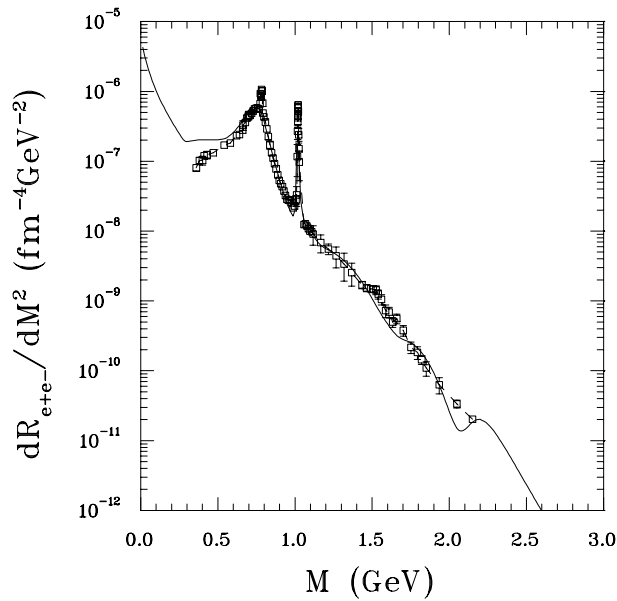}
\caption{Net dilepton production from a gas of mesons at a temperature
of T = 150 MeV, as a function of dilepton invariant mass. The solid
curve is the sum of the hadronic channels discussed in the text and in
the references. The data points are from [\protect\refcite{huang2}].} 
\label{ratecomp}
\end{center}
\end{figure}

Even in a careful analysis of the relevant intermediate invariant mass 
dilepton reactions, some concerns remain. These mainly stem from the
need to account for all the sources of electromagnetic radiation. In
kinetic theory approaches, risks exist of double-counting and possible 
omissions.  In an attempt to bypass those, an approach which allows for a
nonperturbative treatment of the strong interaction and avoids a detailed
enumeration of reactions was devised\cite{huang}. The dilepton emission
rate is interpreted in terms of spectral functions of hadronic currents,
tabulated from low energy $e^+ e^-$ annihilation reactions and from
$\tau$ lepton decays. A differential rate expression, obtained 
in the chiral ($m_\pi \to 0$) limit, reads\cite{huang}
\begin{eqnarray}
\frac{d R}{d M^2} & = & \frac{4 \alpha^2}{2 \pi} M T K_1 (M/T)
\nonumber\\
&\ & \times\,\left[
\rho^{em} (M) - \left(\epsilon - \frac{\epsilon^2}{2}\right)
\left(\rho^V (M) -\rho^A (M)\right) \right]\,,\
\end{eqnarray}
where $T$ is the temperature, $\epsilon = T^2/6 F_\pi^2$, $M$ is the dilepton 
invariant mass, and the superscripts on $\rho$
denote the electromagnetic, vector, and axial spectral functions,
respectively.   These spectral distributions are displayed in 
Fig.~\ref{pwa}.
Using the spectral functions to generate the lepton pair emission rate,
a comparison with the rates obtained via a summation of mesonic reaction
channels is shown in Fig.~\ref{ratecomp}. To summarise, the contributing 
channels producing lepton pairs in the invariant mass range 1 GeV $<$ $M$ $<$
3 GeV have been found to correspond to the initial states $\pi \pi$,
$\pi \rho$, $\pi \omega$, $\eta \rho$, $\rho \rho$, $\pi a_1$, $K
\bar{K}$, $K \bar{K}^*$ + c.c. [\refcite{ioulia}]. 
The detailed channel-by-channel assessment clearly accounts for the
net signal yielded by the ``global'' spectral function analysis. 

In order to compare with experimental data, the rates must be
time-integrated in a model that is also compatible with other measured
observables, hadronic or otherwise. 
Furthermore, a precise simulation of the detector acceptance and
resolution is necessary. An approach that incorporates both aspects is
described presently. A class of models that produce time-evolution
scenarios is that of hydrodynamic models. Specifically, the assumption
is that, at SPS energies, a plasma is produced at proper time $\tau_0$.
Assuming isentropic expansion, the temperature and proper formation time
can be related to the measured differential multiplicity\cite{bj}
\begin{eqnarray}
\frac{2 \pi^4}{45 \zeta (3)} \frac{1}{A_T} \frac{d N}{d y} = 4 a T_0^3
\tau_0\ .
\end{eqnarray}
$d N/d y$ is the measured particle rapidity density and $a = 42.25 \pi^2 /
90$ for a plasma of massless $u$, $d$, $s$, $g$ partons. Once the
transverse area $A_T$ is known along with $d N /d y$, the above relation
links $T_0$ with $\tau_0$.  Enumeration of the model
premises is completed by the statement that the plasma is assumed to
undergo a boost-invariant longitudinal expansion and an
azimuthally-symmetric radial expansion, with a transition to a hot
hadronic gas consisting of {\it all} hadrons having $M$ $<$ 2.5 GeV, in
thermal and chemical equilibrium at temperature $T_c$. This makes for a
rich equation of state. Once all parton matter is converted into
hadronic matter, expansion continues until a kinetic freeze-out
temperature $T_F$ is reached. Those steps are generic in hydrodynamic
calculations. Note that during the evolution, the speed of sound in
matter is consistently calculated at every temperature that is input
into the equation of state and needed to solve the hydrodynamic 
equations\cite{crs}. Additional details about setting up the initial conditions 
for the hydrodynamic evolution can be found in Ref.~[\refcite{kgs}]. The same
reference also shows the result of hadronic spectra calculations with the
hydrodynamic approach.

\begin{figure}
\begin{center}
\includegraphics[width=8cm,angle=0]{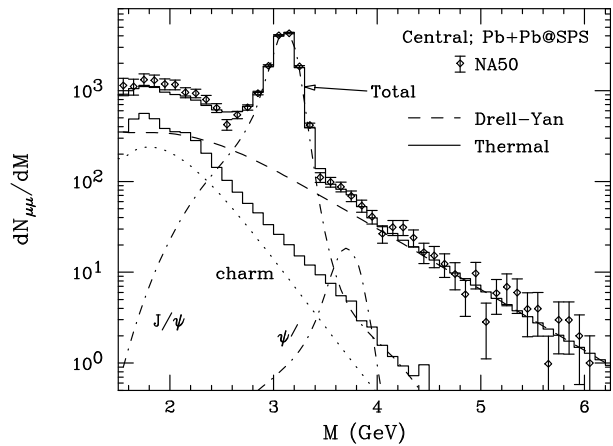}
\caption{The calculated dimuon invariant mass distribution, after
correcting for the detector acceptance and resolution. The data are from
the NA50 collaboration\protect\cite{na50imr}. The Drell-Yan and thermal
contributions are shown separately, as well as those coming from
correlated charm decay and from the direct decays of the $J/\psi$ and
$\psi'$.}
\label{na50M}
\end{center}
\end{figure}

\begin{figure}
\begin{center}
\includegraphics[width=8cm,angle=0]{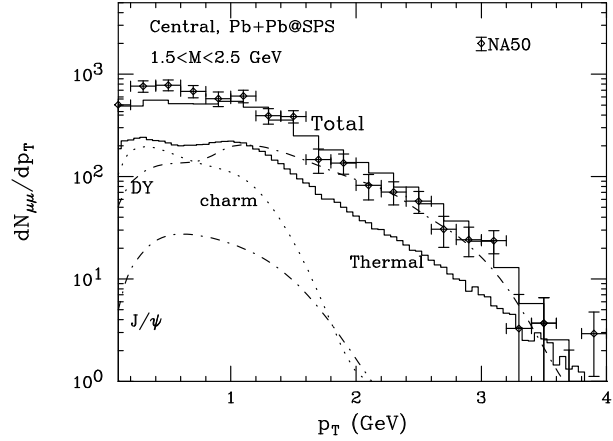}
\caption{The dimuon transverse momentum spectrum, after accounting for
the detector effects. The data and the different curves are from the 
same sources as in Fig.~\protect\ref{na50M}.}
\label{na50pt}
\end{center}
\end{figure}

It is vital to account for the finite acceptance of the detectors and
for their resolution when comparing the results of theoretical
calculations with measured experimental data.  In the case at hand,
those effects are indeed important in the NA50 
experiment\cite{capelli}. One approach to this problem in the past has been to
model approximately and analytically the acceptance\cite{rs,sskg}.
While this can be readily implemented, a legitimate doubt can subsist
about the accuracy of the experimental representation, especially in
regions where edge effects might be important. In order to circumvent
this problem, a numerical subroutine developed to reproduce the NA50
acceptance cuts and finite resolution effects in the measurement  of muon
pairs in Pb + Pb collisions at the CERN SPS was used\cite{kgs}. Thus,
the invariant mass distribution of lepton pairs is computed in the
hydrodynamic model, and then the pairs are run though the numerical
detector simulation. The normalisation is determined by a fit to the
Drell-Yan data using the MRSA parton distribution functions, as in the
NA50 analysis. In order to get a $p_T$ distribution, the $dN /d M^2$
estimates for Drell-Yan were supplemented with a Gaussian distribution in
$p_T$\cite{rs}, this very closely reproduces estimates obtained by the
NA50 collaboration. The resulting invariant mass distribution is shown
in Fig.~\ref{na50M}.
The $p_T$ distribution is also computed. It is shown in
Fig.~\ref{na50pt}. In both cases, good agreement with the experimental data is
clearly achieved.

\begin{figure}
\begin{center}
\includegraphics[width=8cm,angle=0]{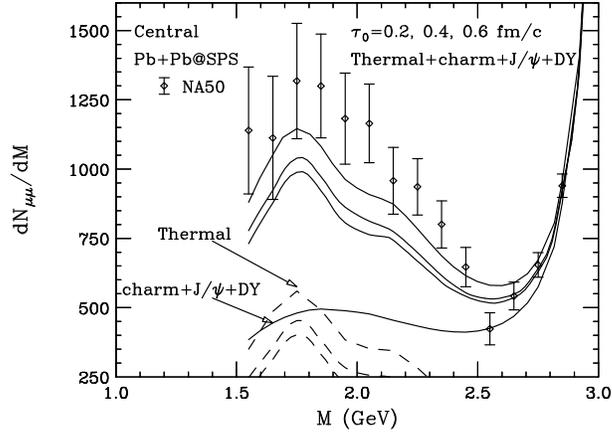}
\caption{A linear plot of the net dilepton spectrum in the intermediate
mass region. The three solid curves correspond to formation time
$\tau_0$= 0.2, 0.4, 0.6 fm/$c$, from top to bottom, respectively. The data
are from [\protect\refcite{na50imr}]. The thermal contribution and that for
hard processes are shown separately. }
\label{na50linear}
\end{center}
\end{figure}

As this point it is appropriate to consider the following question:
which initial temperature is demanded by the intermediate invariant mass
dilepton data? A critical and quantitative assessment of this issue can
be obtained by examining a {\it linear} plot of the lepton pair 
mass spectrum in the region under scrutiny. This is shown in
Fig.~\ref{na50linear}.
From this figure it is clear that the best fit is provided by $\tau_0$ =
0.2 fm/$c$, and that the second best (less than two standard deviations 
away for most of the data points) belongs to $\tau_0$ = 0.4 fm/$c$. In
terms of initial temperatures, those correspond to $T_0 \approx$ 330 and
265 MeV respectively. A conservative and reasonable point of view is
that it is probably not fair in such a challenging and complex
environment as that of ultrarelativistic heavy ion collisions to ask for
an agreement that is better than two standard deviations, considering
all of the inherent uncertainties. The quark matter contribution (as modeled by
$q\bar{q}$ annihilation) is
$\approx$ 23\% for $\tau_0$ = 0.2 fm/$c$, and $\approx$ 19\% for $\tau_0$ = 
0.4 fm/$c$, around a lepton pair invariant mass of 1.5 GeV. 

\begin{figure}
\begin{center}
\includegraphics[width=6cm,angle=90]{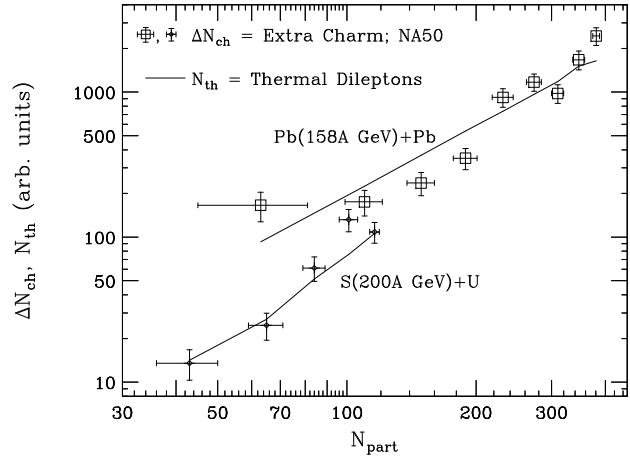}
\caption{Centrality dependence. The data represents the ``extra charm''
yield (as characterised by the NA50 
collaboration\protect\cite{na50imr}) needed to describe the intermediate 
mass dimuon data. The solid curves are from the sources discussed in the 
text.}
\label{na50centra}
\end{center}
\end{figure}

Focus so far has been placed on high multiplicity data only. However, to
extend the hydrodynamic model to non-central events and to properly treat
the azimuthal anisotropy is not a simple task. However, one can get an 
approximate estimate of the centrality dependence by
ignoring the broken azimuthal symmetry and by approximating the region
of nuclear overlap by a circle of radius $R \approx 1.2\, (N_{\rm
part}/2)^{1/3}$, where $N_{\rm part}$ is the number of 
participants\cite{centra,kgs}. A centrality-dependence is generated thusly and
shown in Fig.~\ref{na50centra}. 
It is seen that the agreement with the measured data is quite good and
that this approach gives a fair description of the centrality dependence of
the excess dilepton measurement.

What has been learnt from this exercise is that many-body channels, 
important in the low mass region, also still play a vital role in the
intermediate mass domain. This unity is satisfying. It also happens that the
global dynamical behaviour of the electromagnetic radiation can be empirically 
modeled\cite{kampf}. Perhaps more importantly, the case where the data shown in
this subsection are interpreted as the signature of a charm excess no longer
appears to be very compelling. The findings described here are in agreement with 
previous calculations of dilepton radiation at this mass scale\cite{gali,rs}. 
As mentioned previously, the portion of the signal that emanates from the 
deconfined sector is around 20\%, a figure that is unfortunately too small
to support convincing claims of a QGP presence, once all uncertainties 
are factored in. 

\subsection{Photons}

\subsubsection{General Strategy}

Real photons differ from previously discussed dileptons in a couple
of important ways.  First, they are on-shell and thus cannot be
accommodated kinematically with two-hadron annihilation processes. 
From the beginning then, the only one-loop contributions 
to the current-current correlator (or retarded self-energy) are the 
hadronic radiative decays $\pi^{0}\rightarrow
\gamma\gamma$, $\eta\rightarrow\gamma\gamma$, 
and $\omega\rightarrow\pi^{0}\gamma$.  Although the $\omega$ lifetime
is $\sim$ 23 fm/$c$ and so that channel could be considered a 
thermal source (just barely), the others
are clearly non-thermal sources with lifetimes much longer than the 
fireball.
Such contributions must be considered in the overall yield, but 
are outside of the scope of the present discussion focusing on
thermal emission.
The one-loop contributions therefore play a smaller role in photon
production as compared with dilepton production and consequently the 
discussion begins seriously here at two loops.
Second, and something of a technical point, is that real photons
have only two polarisation states over which to sum, rather 
than three as was the case for virtual photon propagation.   In the same
spirit then as the dilepton case, real photon emission from resonance 
hadronic matter can be most systematically studied by beginning with
the photon self energy at two-loop order and working upward.   The
occupation numbers for the internal lines pay always a Boltzmann penalty 
and it is therefore natural to begin with the lightest species and
with the minimum number of hadrons. As a aside remark, since the evaluation of
two-loop topologies at finite temperature is technically  challenging (although
not impossible), many photon rate calculations rely on a kinetic theory
approach.

Since the energy regime is both relativistic and nonperturbative in
terms of QCD degrees of freedom, it is commonplace to use
effective theories for the composite hadron dynamics.  Typically one
starts with an effective Lagrangian with a large enough flavour symmetry to
account for the lightest and relevant species.  As a general
rule, the pions are most important, followed by rho and so on, simply 
owing to increasing mass.  Quantum numbers also play a role in 
terms of spin states and isospin states governing densities, and 
so one must be systematic.  With interactions
under some control relative to chiral symmetries, gauge invariance,
conservation requirements of various sorts, 
one uses cutting rules on the two-loop
self-energy diagrams in order to generate a list of reactions of the type
$h_{a} + h_{b} \rightarrow h_{1} + \gamma$ and
$h_{a} \rightarrow h_{1} + h_{2} + \gamma$. 

\subsubsection{Establishing the Rates}

The above mentioned strategy was first taken by Kapusta, 
Lichard and Seibert\cite{kls91}
where $\pi$--$\rho$ and light meson dynamics were investigated.  The dynamics were
modeled with
\begin{eqnarray}
{\cal\/L} & = & {1\over\,2}|D_{\mu}\,\Phi|^{2}\,-\,m_{\pi}^{2}\,|\Phi|^{2}
\,-\,{1\over\,4}\rho_{\mu\nu}\,\rho^{\mu\nu}\,+\,{1\over\,2}
m_{\rho}^{2}\rho_{\mu}\rho^{\mu}\,-\,{1\over\,4}F_{\mu\nu}F^{\mu\nu}.\ \
\end{eqnarray}
Coupling of the rho and the photon to pions was accomplished with the
covariant derivative $D_{\mu} = \partial_{\mu} -ieA_{\mu} 
-ig_{\rho}\rho_{\mu}$.  The charged and neutral pions are embodied in
the complex pseudoscalar field $\Phi$, the vector rho and photon
field strength tensors are respectively $\rho_{\mu\nu} = 
\partial_{\mu}\rho_{\nu}-\partial_{\nu}\rho_{\mu}$ and
$F_{\mu\nu} = \partial_{\mu}A_{\nu}-\partial_{\nu}A_{\mu}$.  Calibration
is done by fitting the $\rho\rightarrow\pi^{+}\pi^{-}$ decay rate
with the choice $g_{\rho}$ = 2.9.

The specific channels studied in Ref.~[\refcite{kls91}] were 
dubbed annihilation $\pi^{+}\pi^{-}
\rightarrow\rho\gamma$ and ``Compton scattering'' 
$\pi\rho\rightarrow\pi\gamma$, and finally, neutral rho decay
$\rho\rightarrow\pi^{+}\pi^{-}\gamma$ (essentially the finite
temperature analog of the vacuum process studied by Singer\cite{s63}).
Since the $\eta$ meson mass is intermediate between pion and rho,
its Boltzmann penalty is less than rho's.  Its effects were also 
considered by including the channels $\pi^{+}\pi^{-}\rightarrow\eta\gamma$,
$\pi^{\pm}\rightarrow\pi^{\pm}\gamma$.  Owing mostly to coupling
strengths (or weaknesses), these channels were found to be less
important as compared to the purely $\pi$ and $\rho$ channels by more
than an order of magnitude.   Finally, in this initial
study of photon production, the channel 
$\pi^{+}\pi^{-}\rightarrow\gamma\gamma$ was included, though it
was seen to contribute very little.

The matrix elements for all the processes enumerated above are 
included in Ref.~[\refcite{kls91}] and will therefore not be repeated
here.  The energy dependent invariant rate for producing photons
is then obtained by folding in Bose-Einstein (enhanced, if final) hadron
distribution functions and Lorentz invariant phase space.  For 
instance, for the channels $p_{a} + p_{b}\rightarrow p_{1}+p_{\gamma}$, 
one has
\begin{eqnarray}
E_{\gamma}{dR\over\,d^{3}p_{\gamma}} &=& 
{\cal\/N\,}\int\,|\bar{\cal{\/M\;}}|^{2}(2\pi)^{4}
\delta^{4}\left(p_{a}+p_{b}-p_{1}-p_{\gamma}\right)\,
{d^{3}p_{a}\over(2\pi)^{3}2E_{a}}\,f_{a}\,
\nonumber\\
&\times&
{d^{3}p_{b}\over(2\pi)^{3}2E_{b}}\,f_{b}\,
{d^{3}p_{1}\over(2\pi)^{3}2E_{1}}\,(1+f_{1})\,
{1\over(2\pi)^{3}2},
\label{genrateeq}
\end{eqnarray}
where ${\cal\/N\/}$ is the appropriate degeneracy factor counting
the states.
 
\begin{figure}
\begin{center}
\centerline{\resizebox{5.5cm}{!}{\includegraphics{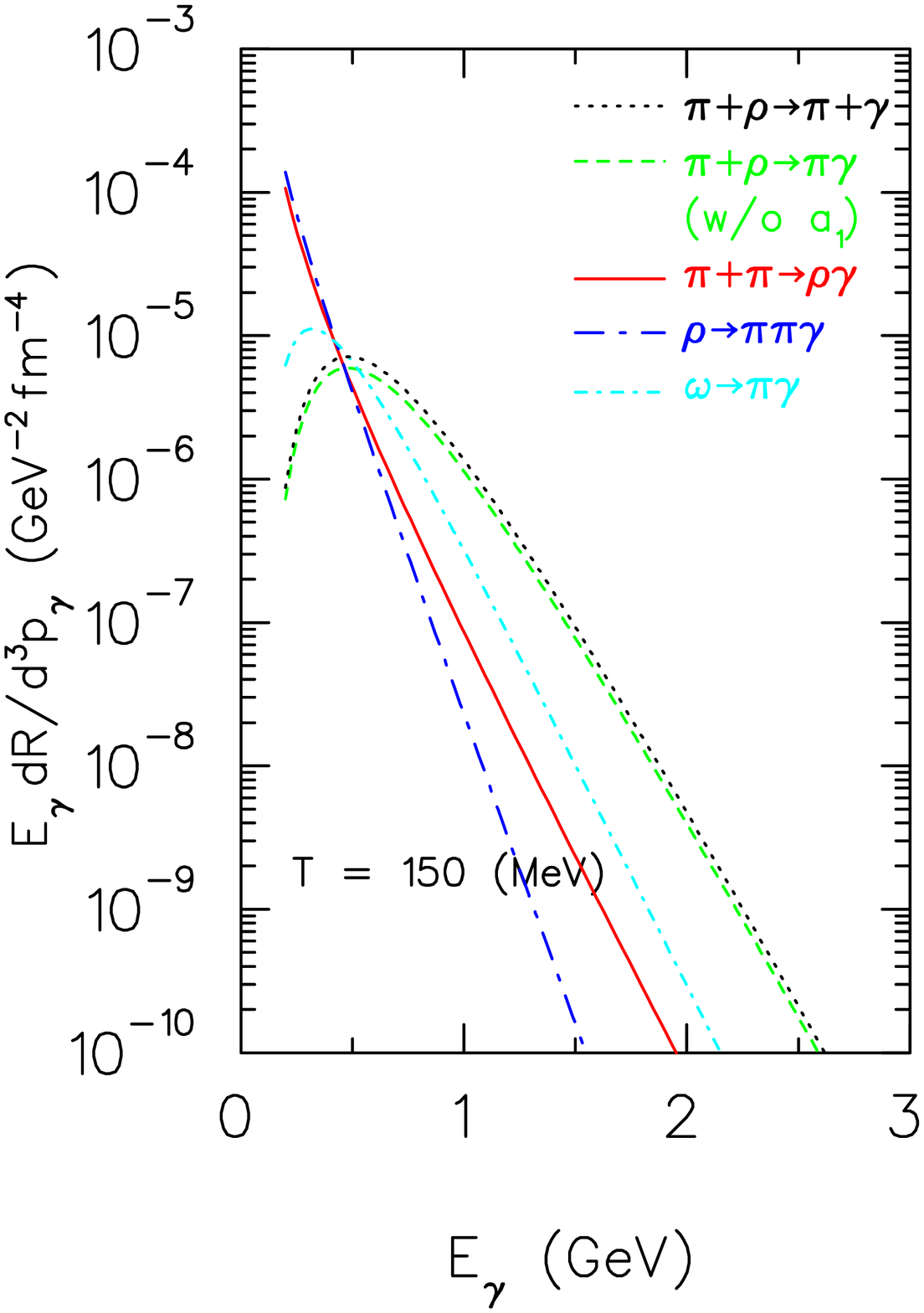}}
\hspace*{0.15cm}\resizebox{5.5cm}{!}{\includegraphics{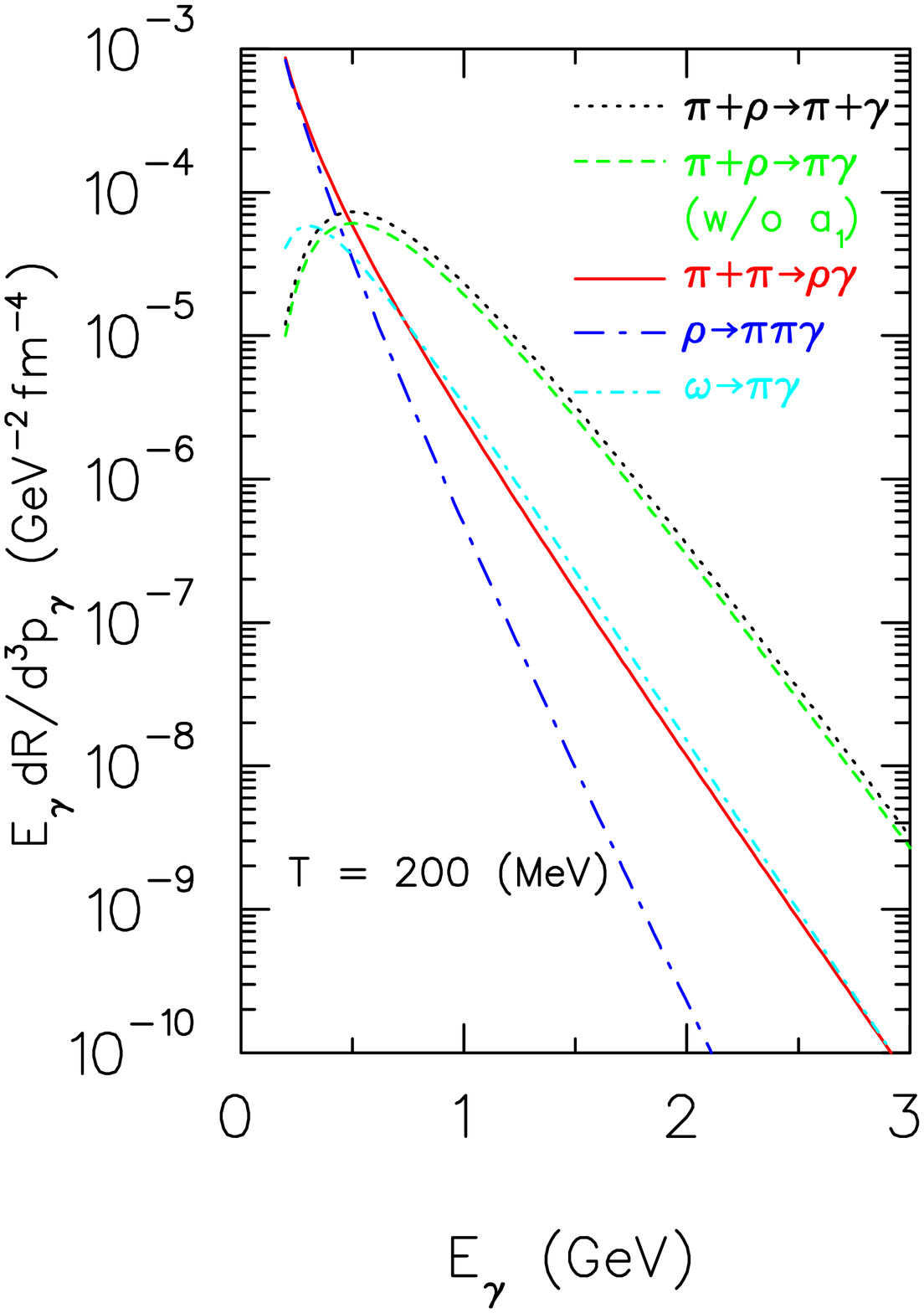}}}
\caption{Photon emission rates from Ref.~[\protect\refcite{nkl92}] plus
an estimate including the $a_{1}$ coherently.  The $\omega\to\pi\gamma$
rate results from Eq.~(\ref{eq:odecay}).  Temperatures
are fixed at T = 150 MeV and T = 200 MeV.}
\label{fig:photonrate1}
\end{center}
\end{figure}

The resulting rates have been established numerically.  However, analytical
parametrisations valid for 100 MeV $<$ $T$ $<$ 200 MeV and 0.2 GeV $<$ 
$E_{\gamma}$ $<$ 3 GeV have been proposed\cite{nkl92}\,\footnote{The
process $\rho\to\pi\pi\gamma$ was slightly miscalculated in 
Ref.~[\refcite{kls91}] 
owing to an omission of a Lorentz-boost factor.  The parametrisation 
of the process published in Ref.~[\refcite{nkl92}] is therefore not optimal.
A slightly different parametrisation is proposed here which
correctly accounts for covariance effects.}.
\begin{eqnarray}
E_{\gamma}{dR\over\,d^{3}p_{\gamma}}({\pi\pi\to\rho\gamma})
&=&  0.0717\,T^{1.866}\,\exp(-0.7315/T\/+\/1.45/\sqrt{E_{\gamma}}-E_{\gamma}/T),
\nonumber\\
E_{\gamma}{dR\over\,d^{3}p_{\gamma}}({\pi\rho\to\pi\gamma}) 
&=&  T^{2.4}\,\exp(-1/(2\,T\,E_{\gamma})^{3/4}\,-E_{\gamma}/T),
\nonumber\\
E_{\gamma}{dR\over\,d^{3}p_{\gamma}}({\rho\to\pi\pi\gamma}) 
&=&  0.1105\,T^{4.283}\,E_{\gamma}^{-3.076+0.0777/T}\,\exp(-1.18\,E_{\gamma}/T).
\label{rates_from_channels}
\end{eqnarray}
In these expressions, $T$ is the temperature, $E_{\gamma}$ is the photon
energy; both must be reported in GeV.  The numerical constants have
appropriate units in each case and the numerical values out in
front have units fm$^{-4}$GeV$^{-2}$.  The numerical results are
shown in Fig.~\ref{fig:photonrate1}.

The rate for the $\omega\to\pi^{0}\gamma$ is also included in 
Fig.~\ref{fig:photonrate1}. 
It can be written as\footnote{Note that
Eq.~(54) of Ref.~[\refcite{kls91}] has an incorrect Lorentz-boost 
factor.  The formula has been corrected here and reported in 
Eq.~(\ref{eq:odecay}).}
\begin{eqnarray}
E_{\gamma}{d\/R\over\,d^{3}p_{\gamma}}(\omega\to\pi^{0}\gamma) & = &
{3m_{\omega}^{2}\,\Gamma\left(\omega\to\pi^{0}\gamma\right)
\over\,16\pi^{3}E_{0}E_{\gamma}}
\,\int\limits_{E_{\rm\scriptstyle\/min}}^{\infty}
d\/E\/_{\omega}\,f_{BE}\left(E_{\omega}\right)
\nonumber\\
& \ &\times\left[1+f_{BE}\left(E_{\omega}-E_{\gamma}\right)\right]
\label{eq:odecay}
\end{eqnarray}
where $E_{\rm\scriptstyle\/min}$ = 
$m_{\omega}(E_{\gamma}^{2}+E_{0}^{2})/2E_{\gamma}\,E_{0}$, and
$E_{0}$ is the photon energy in the rest frame of the
$\omega$ meson.

\subsubsection{Refinements}

Soon after these initial rate calculations were done, Xiong, Shuryak
and Brown\cite{xsb92} pointed out that the $a_{1}$ meson would have an 
important effect on the $\pi\rho\to\pi\gamma$ channel.  It is a resonance
in the $\pi\rho$ sector with pole mass roughly matching the average $\sqrt{s}$
in the fireball and with a rather large width $\sim$ 400 MeV.  An 
interaction for the $a_{1}\pi\rho$ vertex was proposed, stemming not from 
some symmetry argument but rather, a vertex having 
minimal momentum dependence and still respecting gauge 
invariance. The idea was to set the scale as simply as possible
for the contribution from this process.   From the strong-interaction vertex,
vector dominance was employed to subsequently describe radiative decay.  
The $a_{1}$ was indeed found to be important when studied this way---even 
dominating the other exchanges.

However, by itself the $s$-channel $a_{1}$ diagram
does not carry complete information on the overall 
strength of the Compton process.  Instead, a coherent sum of pion exchange
and $a_{1}$ exchange was necessitated.
Song\cite{song93} later carried out a study of photon production using
a Chiral Lagrangian with vector and axial-vector fields
introduced as dynamical gauge bosons of a hidden local symmetry.
The role of the $a_{1}$ once again seemed to be important.  However,
Song found that fitting the relevant masses and coupling strengths
in the model allowed two solutions, {\it\/i.e.\/} two parameter
sets, and was therefore not able to uniquely identify an emission
rate from this channel.  It was later found, while studying
dilepton emission with the same model, that parameter set II was the 
more reliable one in terms of its ability to match the observable hadronic 
quantities\cite{gaogale}.  In particular, the $D/S$ ratio in the scattering 
amplitude for $a_{1}\to\pi\rho$ was more closely 
respected with parameter set II.   One thus observes a diminished presence of
the $a_{1}$ meson (as opposed to estimates cited above) in the
final rate. A modest enhancement of 20\%, attributable to the pseudo-vector,  
is illustrated in Fig.~\ref{fig:photonrate1}. 
An update on this is
forthcoming\cite{rtg03}. A study with the hidden local symmetry approach also
points to a reduced role of the $a_1$ meson\cite{h98}. 

Strange particles have been found to be only marginally important
in the literature up to now.  Specifically, the radiative 
decay $K_{1}\to\,K\gamma$ was considered
and was shown to be strong relative to the non-strange contributions 
only in a limited kinematic domain\cite{kh94}.

A final cautionary remark is in order here.
When the rate spectra are studied at photon energies above 1 GeV, one
must keep in mind that hadronic form factors have not been implemented
in most of the rate calculations up to now (an exception is the
Kapusta, Lichard and Seibert calculation\cite{kls91} which
estimated the form-factor effect on the $\pi\pi\to\rho\gamma$ channel), 
and could result in a suppression of a factor of 2 or more at higher photon 
energies.  This is where the exchanged meson goes further off shell
and brings forward possibly large form-factor effects.  
Advances in this direction will be important.

\subsubsection{Medium Effects}

In terms of higher-order effects, corrections to these rates come in 
at least two forms.  First, there are off-shell effects which can
be conceptualised by dressing
the propagators and vertices for the internal hadron species in the
general photon self-energy structure.  These are the so-called form
factors mentioned earlier in cautionary remark.  Second, and beyond this, 
there
are bona fide medium effects (finite temperature and density
effects, {\it\/e.g.\/} width smearing and pole mass adjustments)
that could be quite important.  The typical pursuit in
studies of medium modifications is to investigate the effects of 
dramatic collision broadened vector meson spectral
distributions\cite{klh95} and/or 
the dropping of the rho mass according to the so-called Brown-Rho 
scaling\cite{br91} or some other ansatz.
Several authors have studied various pieces of the overall 
medium dependences\cite{s98,sf98,r99,h98}.  The trends are the 
following.  While
the in-medium vector meson widths are expected to be rather
large, the effect on photon production is not too
significant.  This makes sense since the vector spectral distributions
contribute to photon production only as an integral over the
specific distribution---and smearing the distribution does
not affect the normalisation.  Mass shifts, on the other hand, have been
shown to affect the rates by anywhere from a factor of 3 up to 
an order of magnitude\cite{s98,sf98}.  The results are too model
dependent to make specific concluding statements at present. 

\subsubsection{Alternative Approach: Chiral Reduction Formulae}

Instead of computing photon production rates using a 
channel-by-channel assessment, Steele, Yamagishi, and Zahed 
used chiral reduction formulae together with a virial expansion
and they came forward with photon and dilepton emission rate estimates.
The general idea is that the invariant production rate is
proportional to the trace over a complete set of 
hadronic states of the hadronic (Boltzmann weighted) 
Hamiltonian convoluted with a current-current correlator.
The hadronic part of the correlator is written as a virial
type expansion truncated in a particular way.  The expansion
coefficients are constrained by various general arguments,
{\em e.g.\/} broken chiral symmetry, unitarity, and gauge invariance
and also, when available, constrained by observed spectra: 
electroproduction, $\tau$ decay, radiative pion
decay, and so on.   The thermal photon emission estimates in this 
approach tend to be larger than those using an effective Lagrangian 
approach by a factor of 2--4\cite{syz96,syz97}.
At present, this might be the honest theoretical error
bar in the rate estimates even after a decade of model calculations. 
Progress continues especially with effective theories in the
hadronic matter converging with results from models firmly rooted
in QCD as the fundamental degrees of freedom as discussed
in the next section.  This is the so-called
duality of hadronic matter and quark matter at the phase boundary
that one expects.

At this stage in the discussion it is somewhat premature to
integrate the photon production rates over a space-time
evolution, which would then facilitate a comparison with experiment,
because radiation from partonic matter has not yet 
been discussed.  So, before considering photon yields from nuclear
collisions and making contact with data, the partonic
contributions to electromagnetic radiation will
be presented, and then yields will be discussed.

\section{Radiation from Partons}

It is of great theoretical importance to establish the production
rates of electromagnetic radiation for many-body systems beyond the
deconfinement phase boundary of nuclear matter.   A model is employed
whereby the matter is assumed to be fully in the partonic phase.  Whilst 
experimental verification of an unequivocal identification of 
thermalised quark-gluon plasma is still forthcoming, it is the 
appropriate picture with which to work as a baseline.  The general formalism
established for photon production rates from hadronic matter 
in Sect.~\ref{impart} is generic to all quantum field theories
and is thus equally valid for partonic degrees of freedom.
And yet, the massless nature of the up and down quarks requires special 
attention. Calculational tools known as hard-thermal-loop (HTL) 
methods have been applied to handle infrared singularities.  Independent
of the experimental advancements then, it would already be important to 
establish quark matter radiative emissivities.
Since heavy-ion experiments at the CERN SPS and at RHIC
have most likely probed into small areas of the deconfined region in
the nuclear matter phase diagram, there is further motivation, and 
indeed some urgency, for theoretical investigations to converge and 
to report emission rates.
Therefore, the status of theory for photon production 
from finite temperature quark matter is discussed below and a separate
section is devoted to dilepton production.

\subsection{Photons}

The imaginary parts of one-loop contributions to the photon 
self-energy, obtained with appropriate cuts, are identically zero due 
to vanishing phase space.
Certain two-loop diagrams give nontrivial contributions.  Cutting rules
provide a bridge between kinetic theory
and field theory where in fact, a mapping has been 
established\cite{aw83}.
The result of cutting two-loop diagrams gives QCD processes
of the types $q\bar{q}\to\/g\gamma$ and  
$qg\to\/q\gamma$ or $\bar{q}g\to\/\bar{q}\gamma$.  These processes, as well
as bremsstrahlung processes, were studied
using perturbative matrix elements two decades ago\cite{es78,kk81,hk85}. 
The results
were unfortunately infrared unstable ({\it i.e.} the rates 
diverged as the quark mass tended to zero).  Significant improvement
came when the ``annihilation'' and ``Compton'' processes were
analysed by Kapusta {\it et al.\/}\cite{kls91} and Baier {\it et
al.}\cite{baier} using  
resummation techniques of Braaten and Pisarski\cite{rp88,bp90}.
The basic idea behind the resummation technique, or the so-called
hard-thermal-loop approximation, is that weak coupling at high
temperatures allows a separation of scales, and a separation
of the rate into soft (quark momentum $\sim\/g\/T\/$ or smaller) and
hard (quark momentum $T$ or larger) contributions.   The soft contribution
can be computed with an appropriately dressed quark propagator in
the one-loop photon self-energy, while the hard contribution
can be computed using perturbative methods and kinetic theory.
In each result, the separation scale appears as a sort of regulator.
When the soft plus hard contributions are collected together and added, the
result is independent of the separation scale, and
of course also independent of quark mass since it was set to zero
from the beginning.  The exact result can be established 
only numerically.  However, using an approximation which is valid for
$E_{\gamma}/T \gg$ 1, a simple pocket formula has been proposed.
At the time, this result was thought to be complete to 
order $\alpha\/\alpha_{s}$.   Specialising to two quark flavours the
result is
\begin{eqnarray}
E_{\gamma}{d\/R\over\/d^{3}\/p_{\gamma}}
& = & {5\over9}{\alpha\alpha_{s}\over\/2\pi^{2}}\/T^{2}\/e^{-E_{\gamma}/T\/}
\ln\left({2.912\over\/g^{2}}{E_{\gamma}\over\/T\/}\right).
\end{eqnarray}
A value of $\alpha_{s\/}$ = 0.4 ($g^{2}$ = 5) is used and
a ``1'' is added to the argument of the logarithm when plotting as suggested
by Kapusta {\it et al.\/} to more closely match the exact
numerical result for photon energies of the order of the temperature.
This also ensures the rate is always positive. 
These results were subsequently generalised to finite quark chemical
potential and also applied to chemical non-equilibrium 
systems.  For a discussion of these effects see [\refcite{mt02}] and
references therein.

Having photon emission rates from QCD free from infrared
instability ailments represented significant advancement and was at the 
time, thought to be the {\em complete\/} lowest order result.  After 
all, the two loop contributions
to the self-energy (that is, the dressed two-loop contributions, which
actually contain arbitrarily many loops) seem naively to contribute
to photon production at ${\cal{O}}(\alpha\alpha_{s}^{2})$.  They
specifically correspond to bremsstrahlung processes and annihilation
with scattering.  The extra vertices would introduce 
an extra power of $g^{2}$ as compared with the one-loop result.  
However, Aurenche {\it et al.\/}
showed that the two-loop HTL contribution is curiously not
of higher order, but instead contributes to order $\alpha\alpha_{s}$
too\cite{pafg98}.   This owes essentially to a collinear singularity
when the exchanged gluon is soft.  The resummed gluon propagator
introduces a $g^{2}$ in its denominator which cancels the ``extra" $g^{2}$
from the additional vertices.  The overall 
contribution to the HTL for this category of two-loop diagrams 
is the same (lowest) order in $\alpha\alpha_{s}$.  
In terms of the kinetic theory analog, these
correspond to such processes as $q\bar{q}\to\/g\bar{q}\gamma$,
$g{q}\to\/g{q}\gamma$, and $q\bar{q}q\to\/q\gamma$,
or $q\bar{q}g\to\/g\gamma$ (and still others with antiquarks).
Two- and even three-loop contributions
were shown to contribute to lowest order, and the rates
continued to rise!

It is fair to say that after these features were pointed out by
Aurenche {\it et al.\/}, the situation appeared to signal a breakdown
in perturbation theory for finite temperature QCD.  
However, it has been shown recently by Arnold, Moore and Yaffe that as long 
as $E_{\gamma}$ $\gg$ $g\/T\/$, there is sufficient
cancellation due to many-body effects so that the lowest-order rate 
is identifiable and fully under control\cite{amy01}. 
This remarkable result represents a significant
advancement in this field.  There were unfinished details
within the topics of bremsstrahlung\cite{pafg98}, magnetic
mass\cite{pafg00} and coherence effects\cite{pafg00b} that
Arnold {\it\/et al.\/} resolved by analysing
multiple-loop ladder
diagrams which introduce multiple scattering interference effects 
of Landau-Pomeranchuk-Migdal (LPM)\cite{lp53,m56}.   A digression
will not be taken here to reproduce the lengthy and specialised
argument, but the result is the following.  The suppression is sufficient 
to regulate the rates because 1-loop, 2-loop and multiple-loop diagrams 
can be consistently resummed to give a finite rate!
The efforts of many people over a decade of work have produced
a complete photon production calculation from QCD to 
lowest order ${\cal{O}}(\alpha\alpha_{s})$.  The 
simple expression below parametrises the exact numerical
solution for two quark flavours.
\begin{eqnarray}
E_{\gamma}{d\/R\over\/d^{3}p_{\gamma}} & = & {5\over\/9}
{\alpha\alpha_{s}\over\/3\/\pi^{2}}T^{2}{1\over\/e^{E_{\gamma}/T}+1}
\nonumber\\
& \ &\times
\left[
\ln\left({3T\over\/g}\right)
+ {1\over\/2}\ln\left({2\/E\/\over\/T}\right)
+ C_{2\to\/2}
+ C_{\rm brem} 
+ C_{\rm annih} 
\right]\,,
\end{eqnarray}
where
\begin{eqnarray}
C_{2\to\/2} & \simeq & 0.041(T/E_{\gamma}) - 0.3615 + 
1.01e^{-1.35E_{\gamma}/T}
\nonumber\\
C_{\rm brem} + C_{\rm annih} & = & 
{0.633\/\ln(12.28+(T/E_{\gamma}))\over(E_{\gamma}/T)^{3/2}}
+ {0.154(E_{\gamma}/T)\over\sqrt{1+(E_{\gamma}/16.27T)}}\,.
\end{eqnarray}

\begin{figure}
\begin{center}
\includegraphics[width=7cm,height=8cm]{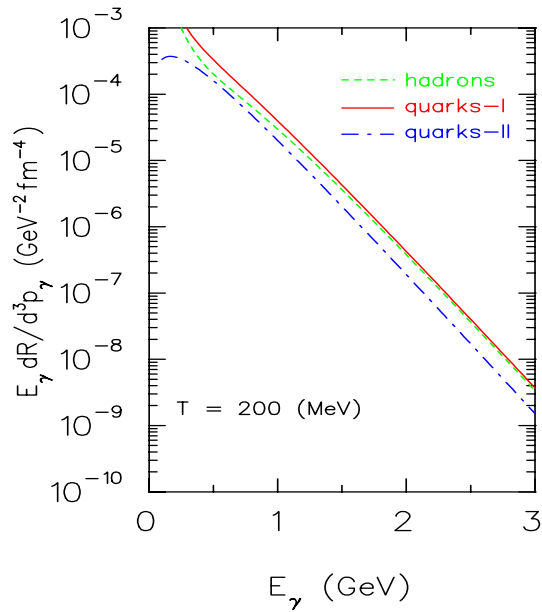}
\caption{Photon emission rate for QGP from Kapusta
{\it et al.\/} in Ref.~[\protect\refcite{kls91}]  
(labelled Quarks--II) and Arnold {\it et al.\/} in 
Ref.~[\protect\refcite{amy01}] 
(labelled Quarks--I).   Quarks--II
includes Compton and annihilation, while Quarks--I includes
in addition, bremsstrahlung and certain 3\/$\to$\/2 processes. 
Quarks--I is the ``complete lowest order calculation''. 
Temperature is fixed at T = 200 MeV.  Total hadron contribution 
is also displayed for comparison purposes.}
\label{quarkrate1}
\end{center}
\end{figure}

Results are shown in Fig.~\ref{quarkrate1}, where
a value for the strong coupling 
$\alpha_{s}$ = 0.4 ($g^{2}$ = 5) is used as before, and superimposed 
onto the total hadron rate discussed previously.
The striking feature is that after 
the dust has settled on the QCD calculations, with HTL to 1-, 2-, and even
multiple-loop order, with LPM effects carefully included, the QCD rate at
fixed temperature is once
again the same as the hot hadronic gas rate.  The QGP and the
hadron gas seem to ``shine just as brightly''.  

\subsubsection{Photon Measurements}

Photon experiments using heavy-ion beams are notoriously
difficult and signals of any kind are already a notable accomplishment.  
At high energies, there are at present two sets of data with which to compare the theory.  
The WA80 collaboration at CERN first reported and discussed their 
yields as absolute measurements, but were later forced to loosen the
constraints somewhat and suggest upper limits only.
Their direct photon limits came from 200{\it\/A\/} GeV $^{32}$S + Au
collisions\cite{wa80_96}.  Secondly, the WA98 collaboration 
measured direct photons in 158{\it\/A\/} GeV $^{208}$Pb + $^{208}$Pb
collisions also at CERN\cite{wa98_00}.  The hope from the onset was
to challenge the theory using production rates convoluted with a temperature
profile evolving according to one of two possible scenarios: 1)
the system first comes into equilibrium well above the phase
boundary and therefore the quark rates contribute until such
time as the system reaches the mixed phase.  Overlapping four
volumes mean that quarks and hadrons contribute until the latent
heat is absorbed fully into a hadronic state, and finally, the
hadrons emit until freezeout; and scenario 2) where the system
reaches a very hot and dense hadronic state and simply radiates
photons while cooling and eventually freezes out.  The burning
question is which scenario is consistent with the measurements?
Can either one be ruled out?

There were several attempts to describe the WA80 results and
do just that.  Shuryak and Xiong\cite{eslx94} first used the hadron rates 
with their version (incoherent treatment) of $a_{1}$
meson dynamics included, and their conclusion was that the excess
photon signal could not be described with a conventional expansion
scenario.  They consequently suggested a long-lived mixed phase
as a possible explanation.  Since the data were later reported
as upper limits only, the conclusion no longer rested on strong experimental
support. 

Srivastava and Sinha applied the quark rates at the 1-loop
HTL level and the hadron rates comparing scenarios (I) with and 
(II) without a phase transition to QGP\cite{ss94}.  They argued that 
the data (which later
became upper limits) are well described by a scenario where QGP
is formed initially.  Bjorken hydrodynamics was employed with $T_{i}$ =
203 MeV, $T_{c}$ = 160 MeV, and $T_{f}$ = 100 MeV for scenario
(I) and, $T_{i}$ = 408 MeV for scenario (II).
Other models came forward attempting to describe
the experimental results.  For example, Dumitru {\it et al.\/} used 
a three-fluid hydrodynamics without and with a phase transition\cite{ad95}. 
They came to similar conclusions, that without a phase transition to
quark matter, the results were inconsistent with experiment.

\begin{figure}
\begin{center}
\centerline{\resizebox{5.55cm}{!}{\includegraphics{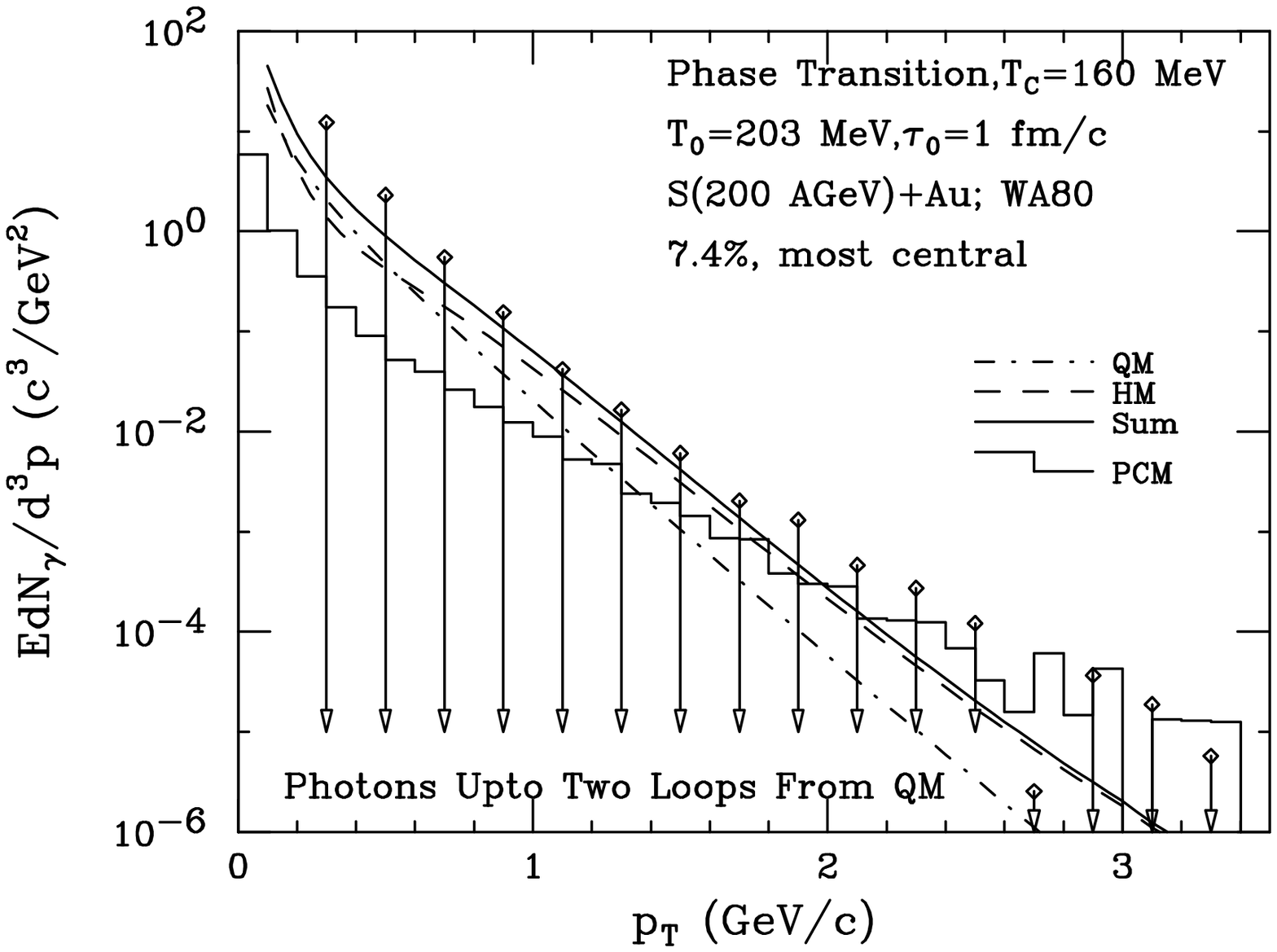}}
\hspace*{0.05cm}\resizebox{5.55cm}{!}{\includegraphics{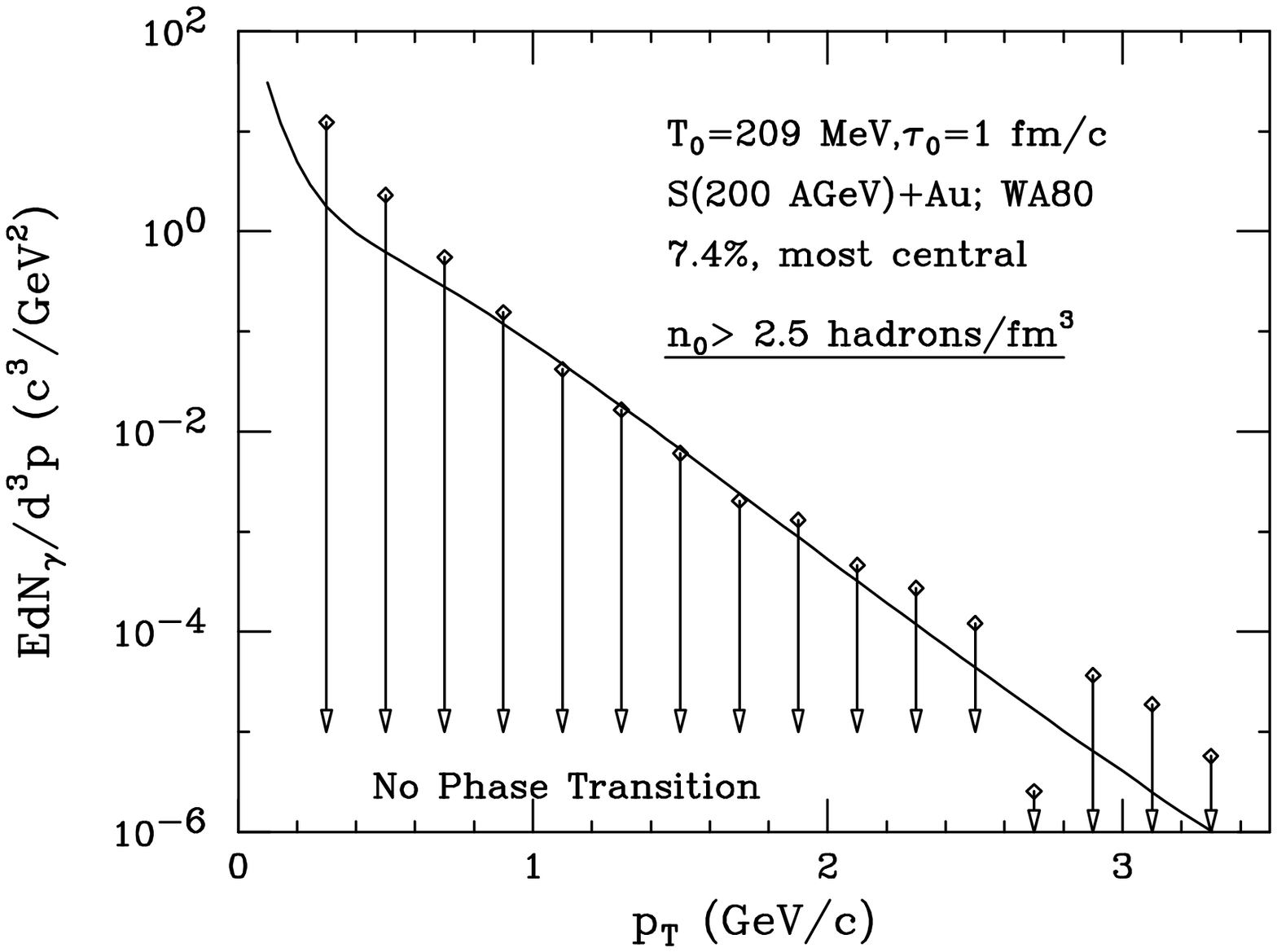}}}
\caption{Upper limits at the 90\% confidence level from 
WA80 on the
invariant excess photon yield per event for the
7.4\% {$\sigma_{\rm\scriptstyle\/mb\/}$} most central 
collisions~\protect\cite{wa80_96} as compared with hydrodynamical
models with (left) and without (right) phase transition}
\label{wa80upperlimit}
\end{center}
\end{figure}

Improvements in rate calculations from quark matter brought
advancement also in yield estimates.  Two-loop HTL rates were
coupled with hydrodynamics, and then later corrected
due to numerical errors along the way\cite{ss00}.  The
most recent and corrected comparison of the WA80 upper limits
to hydrodynamic model estimates are displayed
in Fig.\ref{wa80upperlimit}.  The conclusion is that
both scenarios, without and with a phase transition, seem to
be consistent with the upper limits.  A more complete hadronic equation
of state (EOS) and up-to-date photon rates from quark matter lead to 
these new conclusions.

With the WA80 results in hand and even anticipating the forthcoming WA98
data at that time, Cleymans, Redlich and Srivastava\cite{crs97} used
a hydrodynamical model, which arguably provided better 
description of the evolution as compared to previous model calculations and, 
in particular, could more completely describe the transverse flow likely to 
be generated at the SPS.  The initial QCD rates of Kapusta {\it et al.\/}
were used and the hadron rates, including the effects of the $a_{1}$
meson, were implemented with all hadrons up to 2.5 GeV mass
contributing to the equation of state.  They concluded that while
the final yields were not significantly different in a QCD plus hadron
matter scenario as compared with a fully hadronic picture, they
argued that the physics seemed to favour the former since in the 
hadronic picture particle densities were beyond anything 
reasonable for hadronic language to be justified.

Before moving to the WA98 data, one might make the remark that
since the WA80 results are upper limits rather than measurements, and 
due to the uncertainties in the theoretical production rates and, 
mostly, with the uncertainties in the models for the evolutions of the
nuclear systems, no definite conclusions can be reached.  

The eagerly anticipated direct photon measurement from $^{208}$Pb
+ $^{208}$Pb collisions at 158{\it\/A\/} GeV were published in
2000 by the WA98 collaboration\cite{wa98_00}.  The collaboration
presented their data as compared with several proton-induced reactions
at similar energies and scaled up to central $^{208}$Pb
+ $^{208}$Pb collisions.  For $p_{T}$ $>$ 1.5 GeV, where the signal
is strongest, there is a clear excess beyond that which is expected from
proton-induced reactions.   In other words, the results are
quite suggestive of thermal photon emission, or perhaps pre-equilibrium
emission.  
\begin{figure}
\begin{center}
\includegraphics[width=6cm]{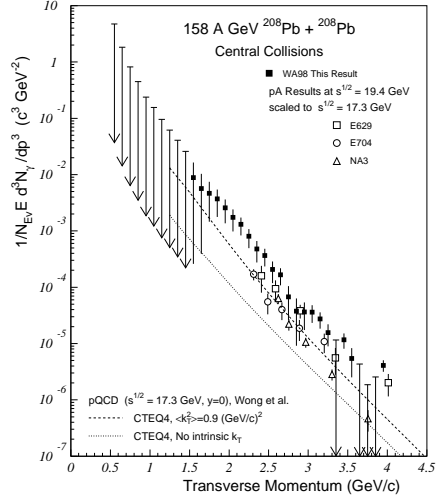}
\caption{The WA98 real photon measurements as a function of photon transverse
momentum. The pQCD estimates are from [\protect\refcite{wong_photon}].}
\label{wa98_pQCD} 
\end{center}
\end{figure}
One contribution that is non-negotiable in those data is that due to
perturbative QCD. It owes its existence to collisions during the first instants
of the reaction, and should appear in pp, pA, and AA measurements. The WA98 data
is shown in Fig.~\ref{wa98_pQCD}, along with a pQCD estimate\cite{wong_photon}.
Even though the presence of pQCD effects at the energies under discussion here 
can't be argued against\cite{pqcd_photon}, the
application to nucleus-nucleus data contains some ambiguities that need to be
pointed out in order to make progress. Specifically, it is clear that some
amount of parton intrinsic transverse momentum (neglected in strict NLO calculations) should 
manifest itself. Simple uncertainty principle arguments 
support this\cite{galeqm01}, and soft gluon emission
should increase the value further\cite{softgl}. However, attempts to extract
meaningful values from experiments have remained inconclusive; for example a recent survey
found that fixed target data at ISR energies ($\sqrt{s}\le$ 23 GeV) were 
inconsistent\cite{auren_kt}. Furthermore, in nucleus-nucleus collisions, a
part of the parton transverse momentum can be ascribed to multiple soft
scattering of the nucleons prior to the hard scattering\cite{bbl}, and this has
to be modeled dynamically and independently. It is
important to note that, at RHIC, several of those uncertainties will be lifted,
as measurements of pp, pA, and AA reactions will be performed {\it at the same
energy} with identical detector configurations. Bearing all those caveats in mind, a
recent study\cite{dumitru_photon} of E704 and WA98 data found that $\langle
k_t^2 \rangle\simeq$ 1.3 GeV$^2$ could by extracted from pp reactions, leaving up
to 1 GeV$^2$ for nuclear effects. This analysis is shown in
\begin{figure}
\begin{center}
\centerline{\resizebox{7.00cm}{!}{\includegraphics[width=8.0cm,height=8.0cm]{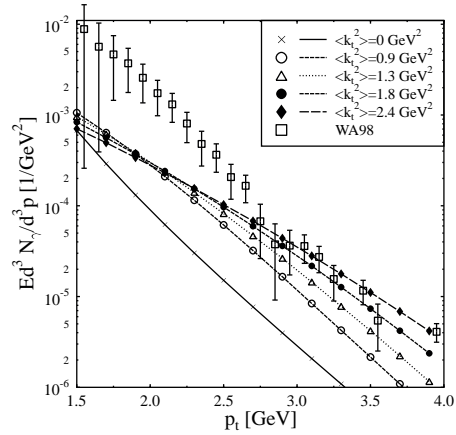}}}
\caption{The WA98 data with pQCD estimates of transverse parton momentum. From
[\protect\refcite{dumitru_photon}].}
\label{dumitru_pho}
\end{center}
\end{figure}
Fig.~\ref{dumitru_pho}.
It is clear from this work that photon transverse momenta below 2.5 GeV are
under-predicted by this pQCD estimate. Also, around this momentum, the exact 
value of the intrinsic
transverse momentum ceases to be important. A softer component of the photon
spectrum is called for, and this will be discussed shortly. Note that this value
for a 
separation of scale  between the ``hard'' and ``soft'' photon sources also
appears if one fits the high momentum pQCD spectrum to the data, with a $K$
factor\cite{gall,kampf}. It is argued in these cited works that the soft
component possesses thermal characteristics. 
\begin{figure}
\begin{center}
\centerline{\resizebox{7.00cm}{!}{\includegraphics{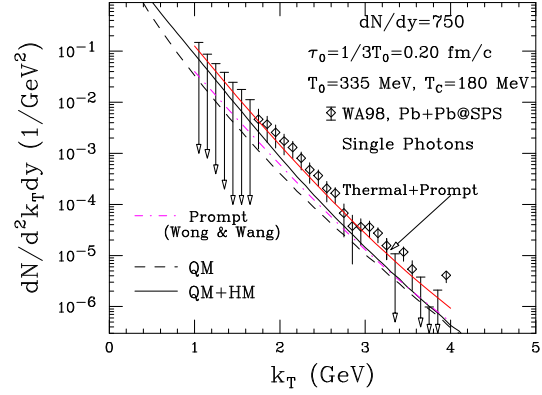}}}
\caption{Single photon production in Pb+Pb collisions at relativistic
energies.  For specific model details, consult 
Ref.~[\protect\refcite{ss01}].}
\label{sswa98}
\end{center}
\end{figure}

\begin{figure}
\begin{center}
\centerline{\resizebox{5.65cm}{!}{\includegraphics{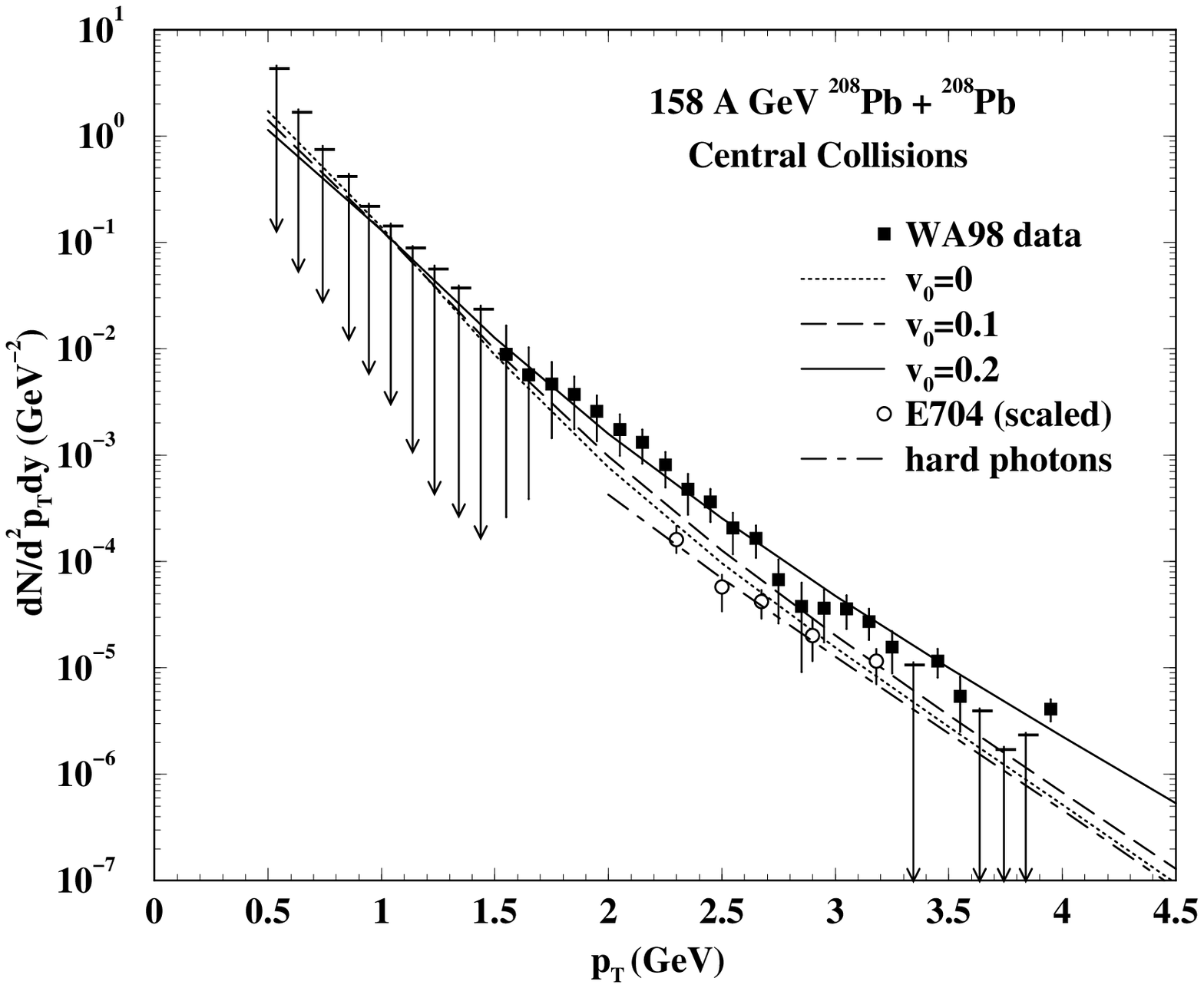}}
\hspace*{-0.025cm}\resizebox{5.65cm}{!}{\includegraphics{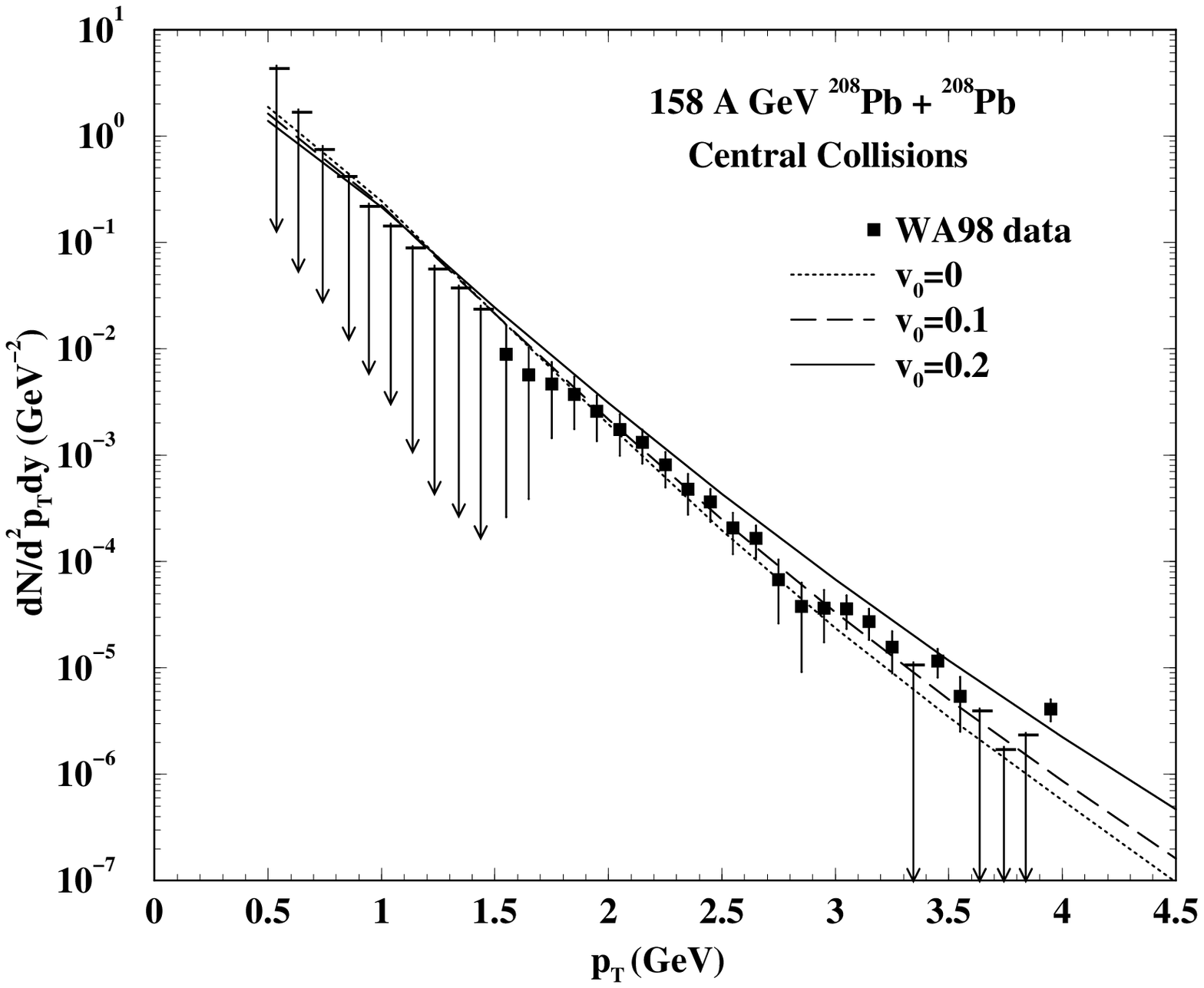}}}
\caption{Total photon yields from quark matter plus hadronic matter
(left panel) and hadronic matter with medium-modified vector meson
properties (right panel).  The figure is reproduced here
from Ref.~[\protect\refcite{jea01}].}
\label{alametalgraphs}
\end{center}
\end{figure}

Srivastava and Sinha\cite{ss01} 
studied mechanisms for excess photon production 
using an hydrodynamic expansion
applied to the $^{208}$Pb+ $^{208}$Pb system.
The photon emission rate from quarks was input using the result
from two-loop HTL calculations from Aurenche {\it\/et al.\/}\cite{pafg98}.  
However, it is probably fair to say that those rates have been superseded by the
calculations in Ref.~[\refcite{amy01}], which incorporates higher loop
topologies and thus LPM effects.   
Production rates from the hadronic phase were taken from the 
parametrisation of Kapusta {\it et al.\/} plus an incoherent $a_{1}$-exchange
contribution to the process $\pi\rho\to\pi\gamma$. The results obtained there
are shown in Fig.~\ref{sswa98}.  
The high initial temperature in this work is needed to generate a sufficient high transverse
momentum component of the photon spectrum. In this respect, the WA98 data has
been used to extract phenomenologically an initial radial velocity
profile\cite{jea01}. The result of that study is shown in
Fig.~\ref{alametalgraphs}. 
Both of those theoretical efforts concluded that 
the excess seemed to be consistent with a
thermal source of photons at roughly $\sim$ 200 MeV temperature,
while detailed and quantitative conclusions on a partonic scenario versus
hadronic with strong flow were not definitively reached.

Ruuskanen and collaborators also used hydrodynamics
to compare theory with experiment\cite{js97,ph99,ph02}.   These 
workers have challenged  
hydrodynamics to find consistency with not only photon spectra, but also 
hadron and dilepton spectra---all within the same model and simultaneously. 
They also insist on reproducing the longitudinal hadron
characteristics\cite{ph02}. 
Several equations of state and therefore several 
expansion scenarios seem to describe the photon spectra and hadron spectra
equally well.  The degeneracy between the different equations of state and
initial conditions is not lifted empirically, even though the data do require
a high density and temperature initial phase. 

\begin{figure}[h!]
\begin{center}
\includegraphics[width=6.50cm]{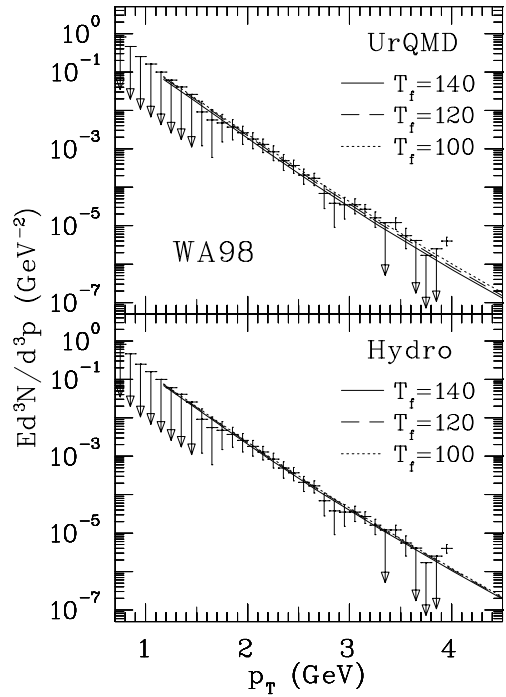}
\caption{Comparison of the WA98 photon spectrum to the predictions
of the UrQMD model and the hydrodynamic model at several freezeout
temperatures from Ref.~[\protect\refcite{hbek02}].}
\label{fig:hydrourqmd}
\end{center}
\end{figure}
A comparison of space-time models in reference to the WA98 data was recently 
carried out
by Huovinen, Belkacem, Ellis and Kapusta\cite{hbek02} wherein hydrodynamics 
and a course-grained UrQMD were used to produce photon spectra (dilepton
spectra were also calculated within theory and compared to experiment). 
Notably, the ``complete'' lowest order photon production rate
[$\cal\/O\/$($\alpha\alpha_{s}$)] from the quark
phase was used in this work.  The basic
conclusions were that UrQMD and hydrodynamics seem to give roughly
the same qualitative features of expansion and cooling (although
quantitatively, UrQMD cools more slowly due presumably to viscous
and heat conduction effects); they give therefore, very similar
results for photon production.  The precise choice of freezeout
temperature seemed to be irrelevant, indicating that the high temperature
part of the evolution dominates photon production.  Results
are shown here in Fig.~\ref{fig:hydrourqmd}.  The agreement
between theory and experiment was described by these authors
as ``excellent'', while they reminded the reader that the rates
have uncertainties and the initial conditions which were fed into the
models are responsible for further uncertainties propagating to the
final spectra.

A partial summary of the photon analyses is justified. It is an accurate
statement that definite conclusions are elusive. Many physical ingredients have
been invoked in the studies of heavy ion photon data, as seen above
and in the quoted reviews, but
uncertainties in many of those ingredients (if not all) preclude a clear interpretation of a
signal that relies on a combination of their effects. But
one example is the absence of the chemical potentials in hydrodynamics-based
approaches. Another is the uncertainty in the basic photon rates. However, those
uncertainties have narrowed down considerably in recent years, and this is true for
rates in both the partonic and confined sectors. Also, as mentioned previously,
the fact of being able to access data at the same energy in pp, pA, and AA
events will make RHIC a fertile testing ground for theoretical models, and
should allow the community to make more progress in differentiating between them.

\subsection{Dileptons}

\begin{figure}
\begin{center}
\includegraphics[width=5.5cm]{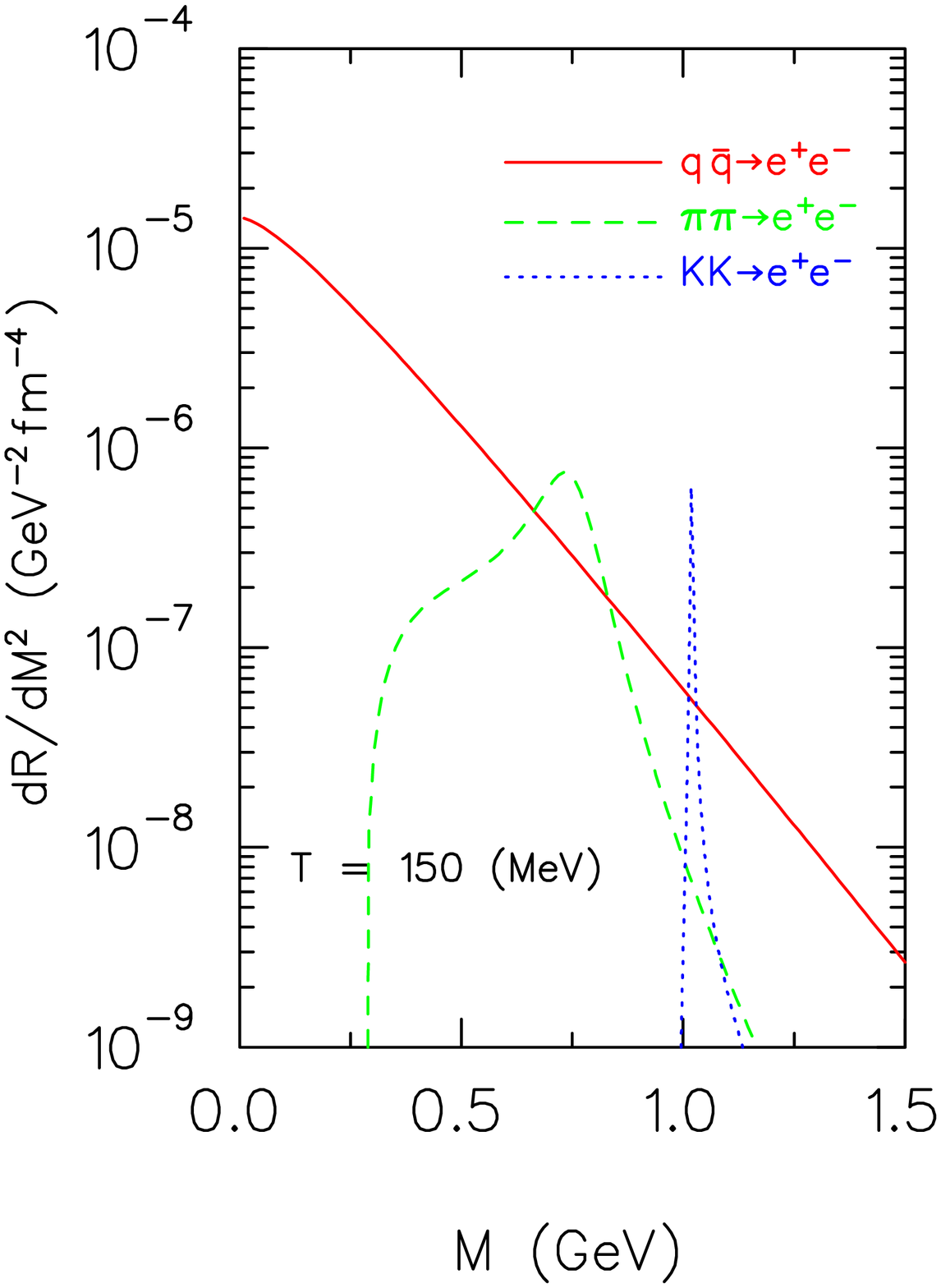}
\includegraphics[width=5.5cm]{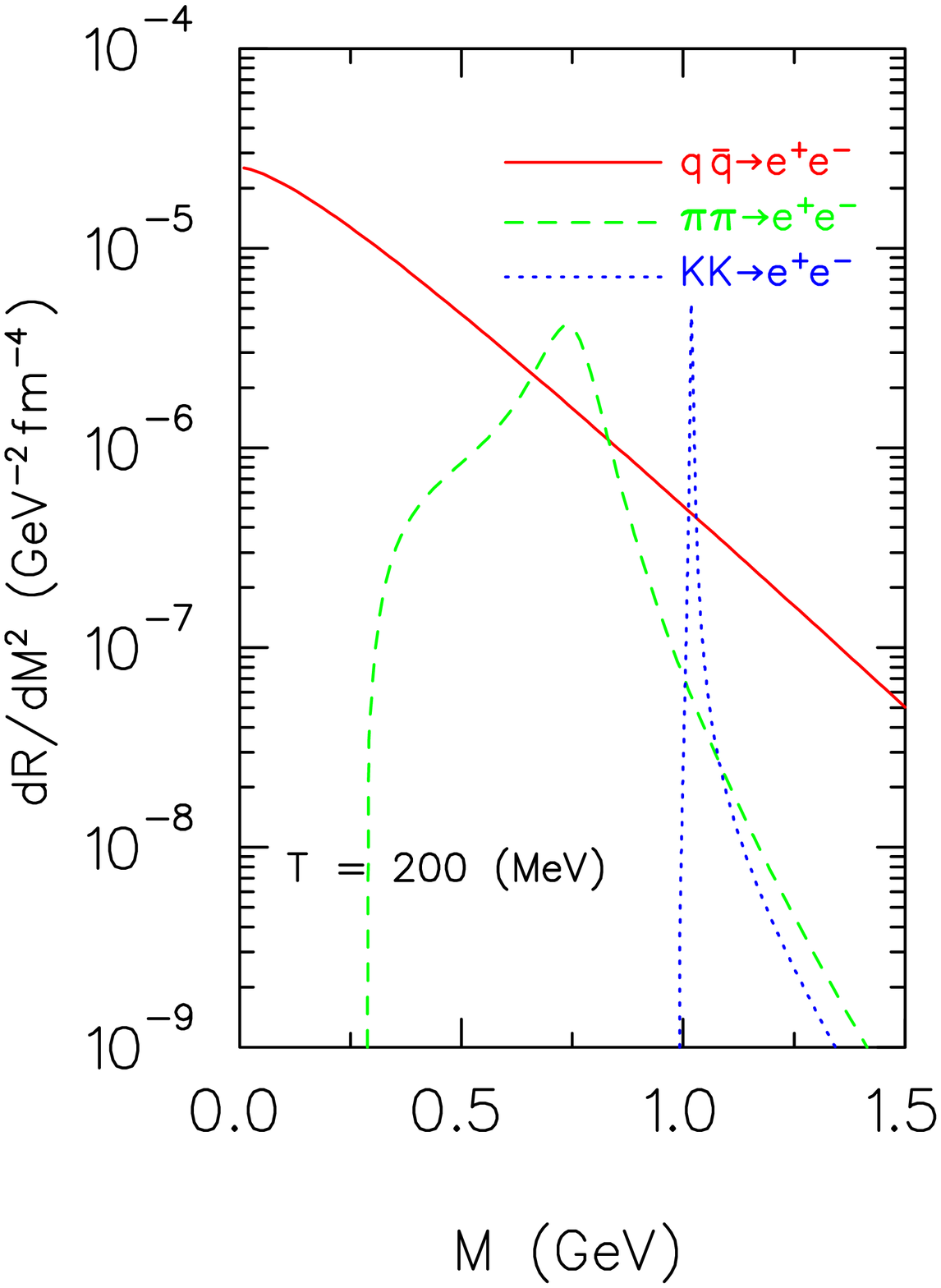}
\caption{Thermal production rate for dileptons via lowest
order quark-antiquark annihilation as compared to leading
hadronic channels $\pi\pi\to\/e^{+}e^{-}$ and
$KK\to\/e^{+}e^{-}$.
The temperatures are set to $T$ = 150 MeV (left) and
200 MeV (right).}
\label{qqbartoee}
\end{center}
\end{figure}

The yield of low mass dileptons ($M$ $<$ $m_{\phi}$) from
thermal quark-antiquark annihilation is not expected to be a
great competitor of the two-pion annihilation simply owing 
to longevity effects in the two phases.  The quark phase occupies a 
smaller four volume.  Nevertheless, it is useful to assess the
production rates as a benchmark
and then to ask about higher-order corrections, especially
in the medium.  For high enough system temperatures, and for large
enough invariant
masses, $q\bar{q}\to\gamma^{*}\to\/e^{+}e^{-}$ is considered
a more significant source and in terms of theory, can be reliably
computed in a HTL approximation.  The lowest order contribution
[${\cal{O}}(\alpha_{s}^{0})$] in field theory language corresponds
to a one-loop graph with bare quarks occupying the internal lines.
The imaginary part of the self-energy describes precisely the annihilation
process mentioned above.  The production rate is roughly the square
of the density of quarks times the cross section times the relative
velocity.  These are now textbook formulae\cite{wong_book} so one simply
quotes the results
\begin{eqnarray}
{d\/R\/\over\/d\/M\/^{2}} & = & {\cal\/N\,}{5\over\/9}{\sigma(M)
\over\/2\/(2\pi)^{4}}M^{3}T\/\,K_{1}(M/T),
\label{dratequarks}
\end{eqnarray}
where the annihilation cross section is
\begin{eqnarray}
\sigma(M) & = & {4\pi\over\/3}{\alpha^{2}\over\/M\/^{2}},
\end{eqnarray}
and where ${\cal\/N\/}$ is an overall degeneracy factor (24 when using
two quark flavours) and finally, $K_{1}$ is the modified Bessel function 
of order 1.  Here the quark and lepton masses have been set to zero.
To compare the resulting rate with major
hadronic contributors, Fig.~\ref{qqbartoee} is presented.  Already here,
one sees that the Born term is not negligible and the natural next
question is the role of higher order contributions.

Perturbative corrections to these annihilation rates were considered by
Braaten, Pisarski and Yuan\cite{bpy90}, who found that for very soft 
dileptons at rest in the medium (energies $\ll$ 1 GeV) the
corrections were orders of magnitude larger than the Born term. 
Also, unique structures emerged in these corrections owing to Van Hove
singularities arising from significant softening of the quark dispersion 
relation in medium.  There appears a minimum in the medium-modified quark 
dispersion relation (a plasmino) typically at dilepton energies less 
than that set by the quark mass.  While these effects are quite intriguing, finite 
imaginary parts in the quark propagators and finite three-momentum effects
for the dilepton\cite{wong} could dampen the peaks into undetectable artifacts.
Also, the softer bremsstrahlung contributions\cite{hge93} might overshine
these total annihilation channels.
For a review of these and other issues for the dilepton
channels, see Ref.~[\refcite{rw00}] by Rapp and Wambach. 

\begin{figure}
\begin{center}
\includegraphics[width=8.0cm,height=8.0cm,angle=-90]{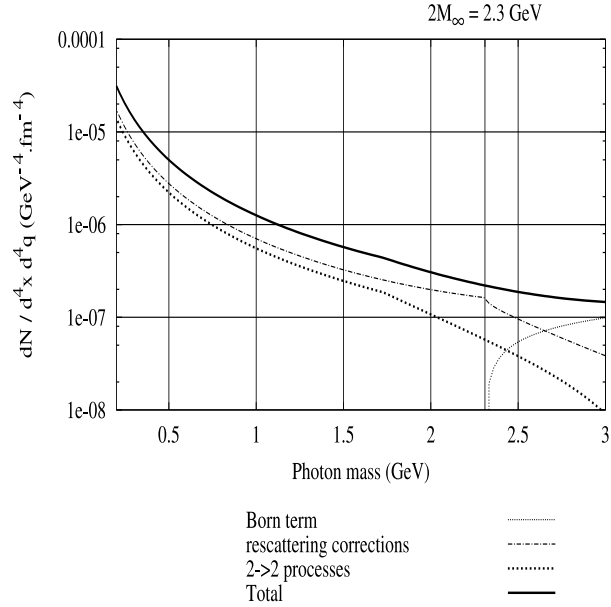}
\caption{The dilepton production rate per unit four momentum at
fixed energy
as a function of photon mass (dilepton mass).   Values
are fixed at $T$ = 1 GeV, $q_{0}$ = 5 GeV, $\alpha_{\rm\/s}$ = 0.3
and two quark flavours were considered.  The figure
is reproduced here from Ref.~[\protect\refcite{agmz02}].}
\label{fig:eefromqgp}
\end{center}
\end{figure}

Quark-antiquark annihilation is of course not the only relevant parton
process for dilepton production.   For instance, the 2$\,\rightarrow\,$2 real
photon production processes considered previously contribute 
also to lepton pair emission. In addition, there
are annihilation processes where one of the incoming partons
has already scattered and suffered an off-shell interaction.
The resulting 3$\,\rightarrow\,$2 process comes from off-shell 
annihilation (also called annihilation with scattering). 
Such mechanisms have been shown to dominate at high enough photon energy.   
Since this is essentially a many-body initial state,
formation time considerations and coherence effects for the virtual
photon suggest once again that multiple scattering plays an important role.
Aurenche, G\'elis, and Zaraket\cite{agz02}, and together with
Moore\cite{agmz02}, have applied the HTL technique for lepton
pairs with $E/T\,\gg$ 1 (either low mass but high momentum, or 
high mass) and have shown that the two-loop contributions which 
include bremsstrahlung of a quark and annihilation with scattering,
are free from infrared and collinear singularity effects.  When 
added to to the Born term, the rescattering corrections plus the
2$\,\rightarrow\,$2 processes\cite{ar92,tt97} result in a rate that 
is somewhat increased as compared with just the Born
terms.  Furthermore, threshold effects at
$M^{2}$ = 4$m_{q\/}^{2}$ (thermal quark mass) are smoothed out,
all of which is illustrated in Fig.~\ref{fig:eefromqgp}.

The dilepton results from QGP discussed above assume equilibrium
and and use asymptotic values for such quantities as thermal
quark masses and screening masses.  Strictly speaking, these are only valid
at asymptotic values of temperature: the assumptions needed for the theoretical
machinery to remain consistent might actually break down at terrestrial 
accelerators energies.
Scenarios more realistic for RHIC and LHC could be
studied if alternative schemes were used to compute masses in nonperturbative
circumstances.  First steps in the direction of lattice evaluations
of thermal dileptons using maximum entropy methods\cite{jg96} have
recently been taken\cite{iw03}.

\section{Predictions}

\subsection{Photons}
RHIC has been running, looking primarily at 
hadronic observables probing the later stages of ultrarelativistic 
nuclear reactions.  Electromagnetic spectra will soon be available which, 
of course, probes deeper into the fireball and indeed cleanly 
into early stages of the reactions as well.  Predictions for measurements of
electromagnetic signals at RHIC are therefore very important.
In addition, the Large Hadron Collider (LHC) is only five years 
away!  While this number probably needs an appropriate error bar,  
it will soon become crucial to have formulated a set of model estimates
for LHC experiments too.  A section is devoted here to discussing these
sorts of predictions.

\begin{figure}
\begin{center}
\includegraphics[width=10.5cm,height=9.5cm]{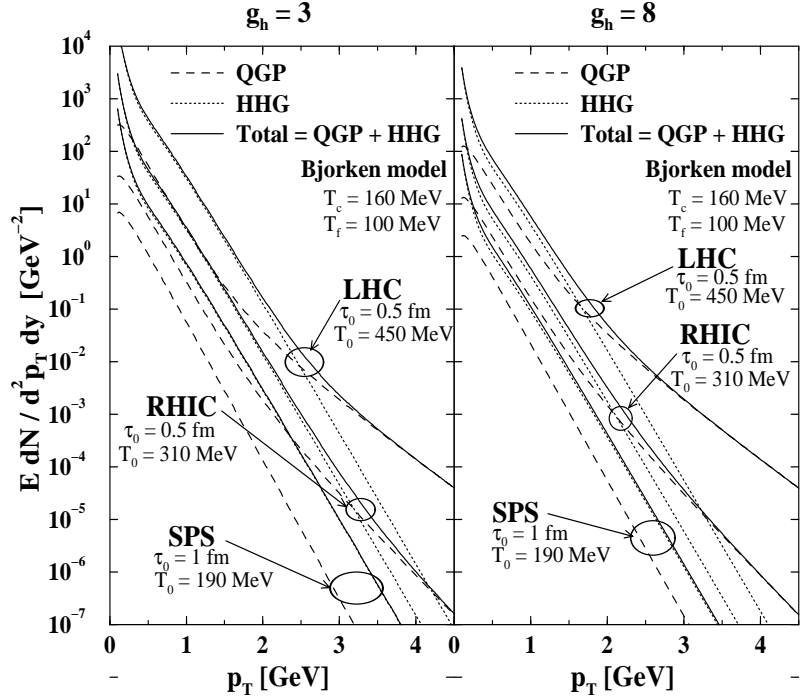}
\caption{Photon spectra (yields) from SPS, to RHIC, and LHC from
Pb+Pb collisions.  Two cases are shown, an ideal pion gas with
$g_{h}$ =3 (left) and 8 (right) panels.}
\label{sps_rhic_lhc}
\end{center}
\end{figure}

As one moves away from SPS systems and energies and goes to RHIC,
and to LHC energies, there are increasing uncertainties 
in estimates for the initial energy densities.  The initial
state is very far from under control.   But, as in all cases, when
theory is extrapolated to new territory, the simplest estimates
are first used to set the scales and subsequent to this, refinements and
various improvements are made.  It is in this spirit that photon
production (yields) were recently estimated at RHIC and LHC
by several authors\cite{jea01,ha98,pp00,st01}.  
Simple 1+1 dimensional models show dominance of the QGP over the 
hadron gas for photon $p_{T}$ ${>}\atop{\sim}$ 3 GeV 
(RHIC) and roughly 2 GeV
(LHC)\cite{st01}.  Transverse expansion, which builds up particularly
later in the hadron phase, makes distinction less clear, but
the QGP might still outshine the hadron gas.  The
results for photon production from Ref.~[\refcite{st01}] are 
displayed in Fig.~\ref{sps_rhic_lhc}.

However, agreement is far from complete on this issue.
For instance, Hammon {\it\/et al.\/}\cite{ha98} predict that QGP will not 
be visible at RHIC owing to a very strong contribution from prompt 
photons (a pre-equilibrium source which has not been discussed here, and
one which is probably not entirely under control), while at LHC the 
situation is different where QGP will be 
visible for limited photon kinematics.  They used 
$T_{0}$ = 533 MeV (300 MeV) and $\tau_{0}$ = 0.12 fm/$c$ (0.5 fm/$c$)
for QGP (hadron gas) at RHIC, and they used
$T_{0}$ = 880 MeV (650 MeV) and $\tau_{0}$ = 0.1 fm/$c$ (0.25 fm/$c$)
for QGP (hadron gas) at LHC.  One could however hope that the
prompt photons due to pQCD could be measured separately 
(in pp collisions at the same energy, for example), and subtracted
out.
Alam {\it et al.\/}\cite{jea01} find, using less extreme initial
condition parameters (lower initial temperatures), that thermal 
photons will be visible for $p_{T}$ $<$ 2 GeV.  However, they
also find that thermal photons from hot hadronic gas populate the high $p_{T}$
region even fairly strongly.  Again, this 
is due to a strong flow built up later in the hadron phase.

In the face of such lack of agreement, which owes essentially
to large uncertainties in the initial conditions, in the
nature of the expansion, and even in the quark and hadron
rates themselves, one suggests that it is premature to make any
definite statement at this point.  In other words, the
theoretical error bar is too large at present to formulate
any physics conclusions from photons.  And yet on the optimistic
side, theory will progress
when the newest QCD rates and hadronic rates are implemented into a 
dynamical model which attempts to describe the buildup of collective 
flow in some detail, on a species-by-species basis
(viz. heavier species seem to flow differently from lighter ones).

\begin{figure}[h!]
\begin{center}
\includegraphics[width=7.0cm]{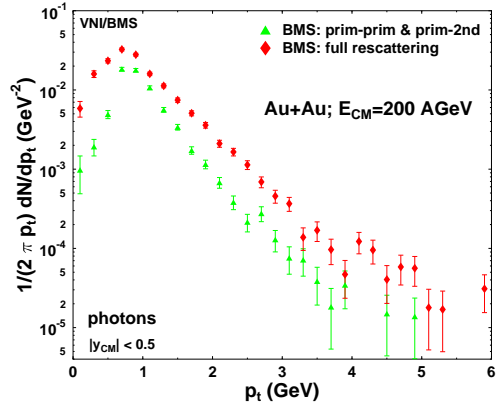}
\caption{Transverse momentum spectrum of photons from central collisions of gold
nuclei at RHIC, calculated with the VNI/BMS parton cascade
model\protect\cite{bms} (see the text for an explanation of the symbols). }
\label{BMS_photons}
\end{center}
\end{figure}

In almost all cases discussed in this work, the dynamical simulations
used to model the dynamics of nuclear collisions assume some form of
equilibrium. Many approaches assume both chemical and thermal equilibrium, while
some only need the latter ingredient. There exists, however, a whole class of
models that attempt an {\it ab initio} rendering of the heavy ion reactions.
Those are currently the only window one has to the very early stages of the
collisions, and thus they potentially offer precious insight on the importance
of pre-equilibrium generation of electromagnetic radiation. At ultrarelativistic
energies, the degrees of freedom that appropriately describe this phase are
partonic. A recent prediction of the photon yields has been
made\cite{bms_photons}, using a version of the parton cascade model
(PCM)\cite{pcm,bms}. Along
with gauging the importance of the above-mentioned pre-equilibrium effects, this
calculation involves the application of perturbative QCD (pQCD) in a domain not
necessarily restricted to large momentum transfers. The photons there are
produced from Compton, annihilation, and bremsstrahlung processes at the parton
level. All lowest-order QCD scatterings between massless quarks and gluons are
included in this model. The obtained photon spectrum for the collision of gold
nuclei at RHIC is shown in Fig.~\ref{BMS_photons}.
There, the contribution from interactions involving at least one primary 
parton (triangles) is compared with that obtained with full binary cascading
(diamonds). Most of the photons between 2 and 4 GeV have their origin in the
multiple semi-hard scattering of partons. This finding would support the claim that
high energy quarks going through a quark gluon plasma would yield
electromagnetic signatures (see later sections).
\subsection{Dileptons}

\begin{figure}
\begin{center}
\includegraphics[width=7.0cm,angle=-90]{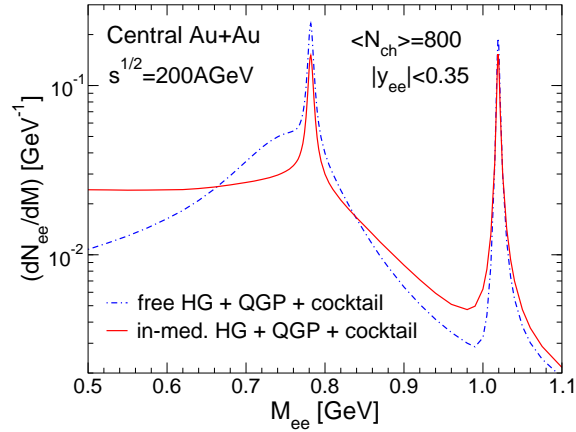}
\caption{Net dilepton yield from an initial plasma phase evolving into a final hadronic 
phase. The full and dash-dotted lines respectively represent cases with and without
in-medium modifications of the vector meson spectral densities ($\rho$,
$\omega$, and $\phi$) [\protect\refcite{rhic_dilep}]. The ``cocktail''
contribution represents decays from on-shell vector mesons, at the end of the
hadronic evolution.}
\label{fig_rhic_dilep}
\end{center}
\end{figure}

The plasma signature in the lepton pair channels is expected to manifest itself 
mainly in the so-called intermediate mass sector\cite{shu78} (see Section
(\ref{interm_sect})). Owing to the large
multiplicities germane to the collider conditions, a large background will
render the extraction of any direct electromagnetic signal from the low mass
region prohibitively difficult. However, there may be still hope to observe some
distortions of the vector meson spectral densities. Using a dynamical simulation
that accounts for a possible under-saturation of the parton chemical
abundances, and estimates of the vector self-energies in a finite temperature
meson gas, the yield on low invariant mass lepton pairs was calculated in
Ref.~[\refcite{rhic_dilep}] and is shown in Fig.~\ref{fig_rhic_dilep}. 
It can be seen that the in-medium effects translate into a suppression of the
$\rho-\omega$ complex, and an enhancement below $M$ = 0.65 GeV and above $M$ =
0.85 GeV. The broadening of the $\omega$ is a candidate for experimental
observation. 

\begin{figure}
\begin{center}
\includegraphics[width=7.0cm,angle=-90]{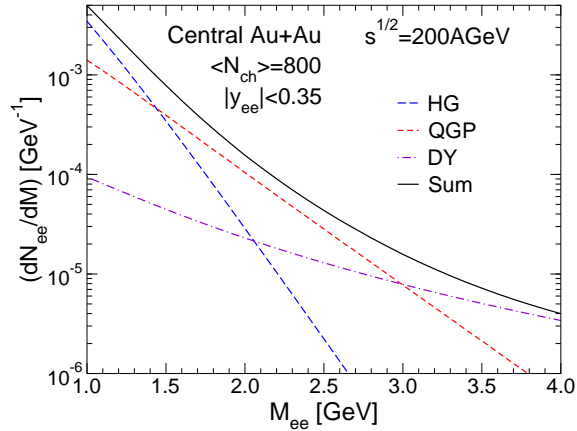}
\caption{Intermediate-mass lepton pair spectra at RHIC energies
[\protect\refcite{rhic_dilep}]. Contributions from the hadron gas (HG),
quark-gluon plasma (QGP), and Drell-Yan (DY) are shown individually, along with
their sum.} 
\label{fig_rhic_dilep2}
\end{center}
\end{figure}

Moving to the intermediate mass region, one obtains\cite{rhic_dilep} the results displayed in
Fig.~\ref{fig_rhic_dilep2}.
In this calculation, the sensitivity of the results on parton equilibrium has 
been examined, and the reader is invited to consult the relevant reference
for the details. The bottom line, however, is that the quark-gluon plasma
contribution (as approximated by Born-term $q \bar{q}$ annihilation) does {\it
not} dip below the Drell-Yan. This being said, the differential invariant mass
distribution is predicted to be
dominated by the semileptonic decays of correlated $c \bar{c}$
pairs, in the intermediate mass region\cite{ramona}.  This  source has not been
shown in the figures above. However, if the heavy quarks that are progenitor for the
semileptonic decay loose energy in the strongly interacting medium, this background 
will be suppressed\cite{heavyQ}. A direct measurement would go a long way in
lifting the ambiguities\cite{na60}.

\subsection{Electromagnetic Signatures of Jets}

It is appropriate to discuss the electromagnetic signatures of jets in an 
environment devoted to predictions, as the physics necessary for those to exist
necessitates high energy and intensity machines such as RHIC and the LHC. The
fate of high energy jets traversing hot and dense matter is a fascinating study,
and this whole subfield has become known as that of ``jet quenching''. The
manner in which high energy jets loose energy in a strongly interacting medium
has been shown to depend on the nature of the medium itself\cite{jets}. 
Thus, jet tomography is expected to be a sensitive probe of the quark-gluon
plasma. However, if jets and plasma interact in such a way that the jet
characteristics are modified, the jet-plasma interactions could by the same
token lead to the emission of electromagnetic radiation. As discussed earlier,
the microscopic processes leading to real photon emission at the parton level
are quark-antiquark annihilation, Compton scattering, as well as bremsstrahlung.
Therefore, a fast quark passing through the plasma will produce photons  by
Compton scattering with the thermal gluons and annihilation with the thermal
antiquarks\cite{FMS}. Those processes are higher-order in $\alpha_s$, when compared with
photons from initial hard scatterings, but they will not form a sub-leading
contribution as they correspond to multiple scattering (actually, double
scattering) which grows with the system size. 

\begin{figure}[t!]
  \begin{center}  
  \includegraphics[width=7.6cm]{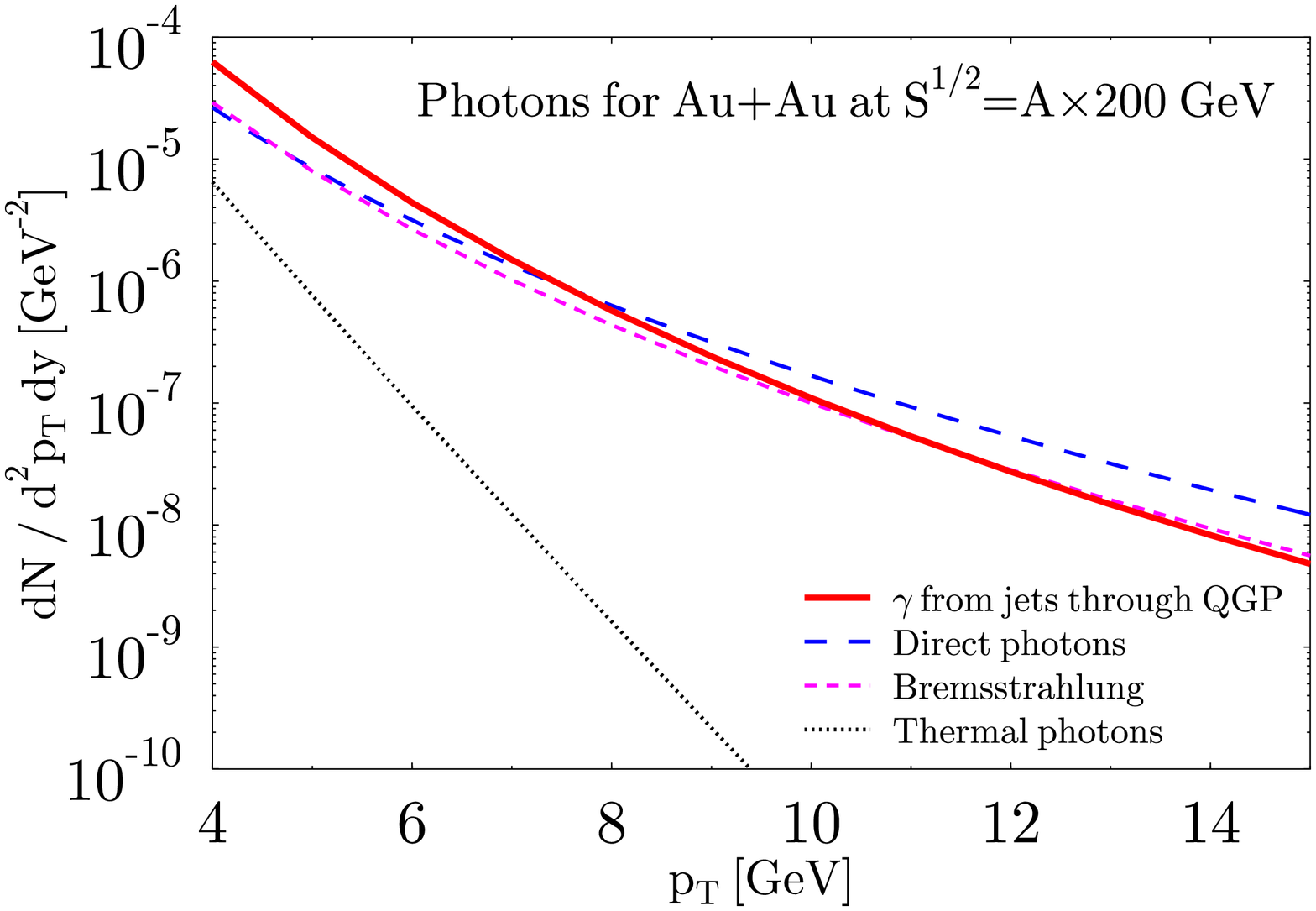}
  \caption{Spectrum $dN/ d^2p_\perp dy$ of photons at $y=0$ for central
   collision of gold nuclei at $\sqrt{S_{NN}}=200$ GeV at RHIC. 
   Plotted\protect\cite{FMS} is the yield for photons
   from jets interacting with the medium (solid line), direct hard photons
   (long dashed), bremsstrahlung photons (short dashed) and 
   thermal photons (dotted).}
  \label{fig:photonrhic}
  \end{center}
\end{figure}

\begin{figure}[t!]
  \begin{center}  
  \includegraphics[width=7.6cm]{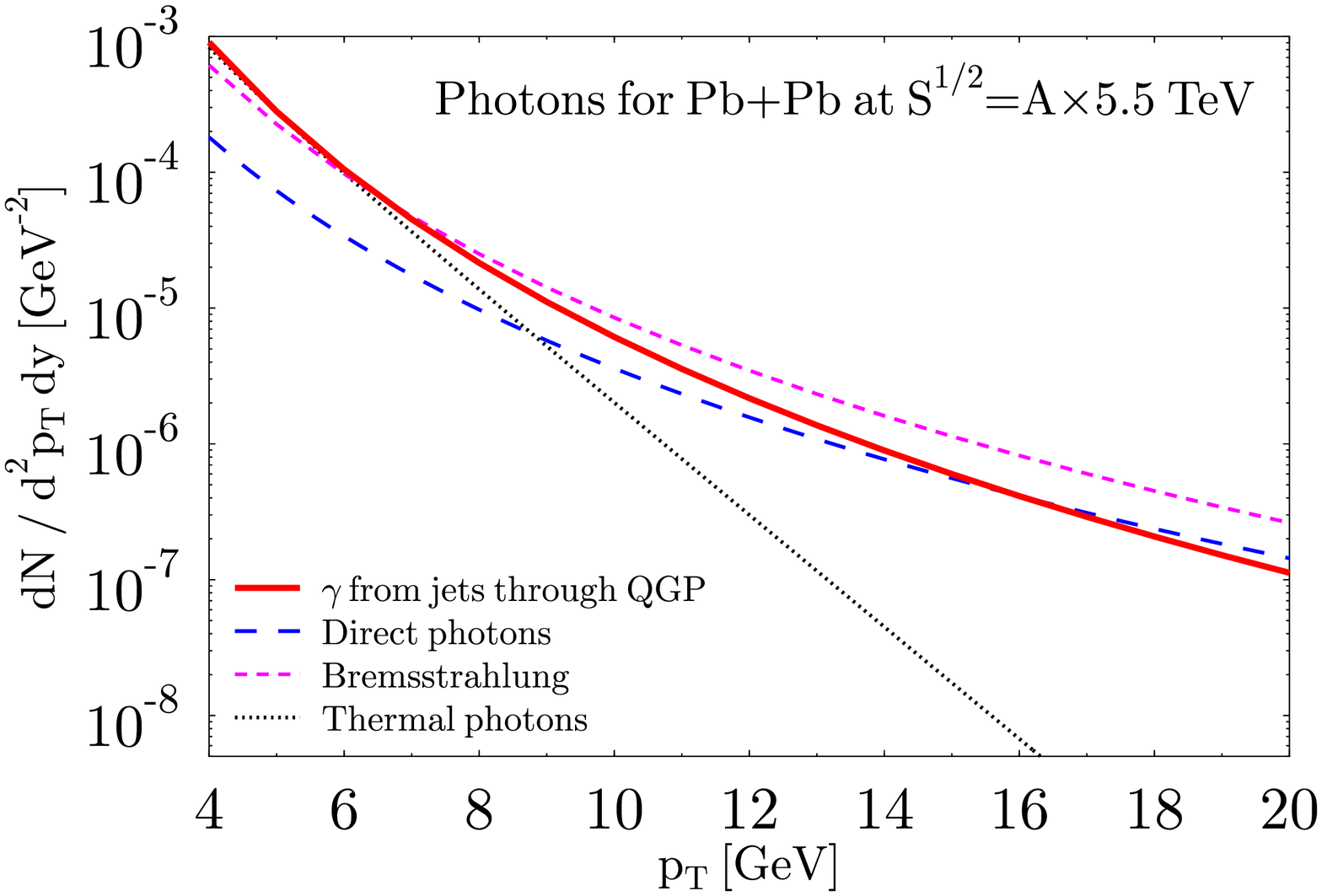}
  \caption{The same as Fig.~\ref{fig:photonrhic} for central collision
  of lead nuclei at $\sqrt{S_{NN}}=5.5$ TeV at LHC. Taken from
  the same reference.}
  \label{fig:photonlhc}
  \end{center}
\end{figure}

Working out the details, one can show that the rate of production of
real photons due to annihilation and Compton scattering is\cite{wong_book}
\ba
  E_{\gamma}\frac{dN^{\rm (a)}}{d^4x \, d^3p_{\gamma}}
  &=&\frac{16 E_\gamma}{2(2\pi)^6}
  \sum_{q=1}^{N_f} f_q({\bf p}_\gamma) \int d^3p
  f_{\bar q}({\bf p}) \left[ 1+f_g({\bf p})\right] \nonumber \\
  & & \times 
  \sigma^{\rm (a)}(s)
  \frac{\sqrt{s(s-4m^2)}}{2E_\gamma E}
  + (q \leftrightarrow \overline{q}) \>,
  \label{eq:ann}  \\
  E_{\gamma}\frac{dN^{\rm (C)}}{d^4x \, d^3p_{\gamma}}
  &=& \frac{16 E_\gamma}{(2\pi)^6} 
  \sum_{q=1}^{N_f} f_q({\bf p}_\gamma) \int d^3p
  f_g({\bf p})   \left[ 1-f_q({\bf p})\right] \nonumber \\
  & &  \times 
  \sigma^{\rm (C)}(s)
  \frac{(s-m^2)}{2E E_\gamma } 
  + (q \rightarrow \overline{q}) \>.
  \label{eq:comp}
\ea
The $f_i$ are parton distribution functions. In order to proceed one may assume
that those may be decomposed as 
\ba
f(\mbox{\boldmath $p$}) = f_{\rm thermal}(\mbox{\boldmath $p$}) + f_{\rm jet} (\mbox{\boldmath $p$}) 
\ea
where the thermal component is characterised by a temperature $T$: $f_{\rm
thermal} = \exp(-E/T)$.  This separation is kinematically reasonable as the jet
spectra fall of as a power law and can thus easily be differentiated from their
thermal counterpart. The phase space distribution for the quark jets propagating
through the QGP is given by the perturbative QCD result for the 
jet yield\cite{lingyu}:
\ba
f_{\rm jet}({\bf p})&=&\frac{1}{g_q}\frac{(2\pi)^3}
  {\pi R_\perp^2 \tau p_\perp} 
  \frac{dN_{\rm jet}}{d^2p_\perp dy} \> R(r) \\& &   \times
  \delta (\eta-y) \Theta(\tau-\tau_i) \Theta (\tau_{\rm max} - \tau) 
  \Theta(R_\perp - r)  \>,
\label{fjet}
\ea
where $g_q=2\times 3$ is the spin and colour degeneracy of the
quarks, $R_\perp$ is the transverse dimension of the system, 
$\tau_i \sim 1/p_\perp$ is the formation time for the jet and $\eta$ is the 
space-time rapidity. $R(r)$ is a transverse profile function.
$\tau_{\rm max}$ is the smaller of the life-time $\tau_f$ of 
the QGP and the time $\tau_d$ taken by the jet produced at position $\bf r$ 
to reach the surface of the plasma. The boost invariant correlation 
between the rapidity $y$ and $\eta$ is assumed\cite{bj}. 
Fig.~\ref{fig:photonrhic} contains the results for thermal photons, 
direct photons due to primary processes, bremsstrahlung photons and the 
photons coming from jets passing though the QGP in central collision
of gold nuclei at RHIC energies. The corresponding results for LHC energies 
are shown in Fig.~\ref{fig:photonlhc}. It is seen that the quark jets passing 
through the QGP give rise to a large yield of high energy photons. 
This contribution should be absent in $pp$ collisions. 
For RHIC this contribution is the dominant source of photons
up to $p_\perp \approx 6$ GeV. The jet-to-photon 
conversion falls more rapidly with $p_\perp$ than the direct 
photon yield, similar to a higher twist correction. 
It is clear that this new mechanism for the production of high energy photons
contributes significantly. In fact, it is the leading source of directly
produced photons at RHIC in the region $p_{\rm\/T} \leq 6$ GeV/$c$.

Very similar considerations should apply to the production of lepton pairs, even
though the details will of course change. Phase space now allows the direct
annihilation of a quark and an antiquark into a dilepton. The cross section for
this process is 
\ba
  \sigma(M^2)= \frac{4\pi}{3}\,\frac{\alpha^2}{M^2}\,N_c (2s+1)^2 \, 
  \sum_f e_f^2\,,
\ea
where the sum runs over the flavour of quarks, $N_c=3$, and $s$ and $e_f$
stand for the spin and the charge of the quark. 
Using kinetic theory, the reaction rate for the above process can 
be written as
\begin{equation}
R=\int \frac{d^3p_a}{(2\pi)^3} f_a({\bf{p}}_a) \int \frac{d^3p_b}{2\pi^3}
  f_b({\bf{p}}_b) \sigma(M^2) v_{\rm rel} \>,
\end{equation}
where $f_i$ stands for the phase-space distribution of the quark or the
antiquark, ${\bf{p}}_a$ and ${\bf{p}}_b$ are their momenta respectively 
and the relative velocity is (for massless quarks)
\begin{equation}
v_{\rm rel}=\frac{E_a E_b -{\bf{p}}_a \cdot {\bf{p}}_b}{E_a E_b} \>.
\end{equation}

After some algebra this can be rewritten as
\ba
\frac{dR}{dM^2}&=&\frac{M^6}{2}\frac{\sigma(M^2)}{(2\pi)^6} \int
\tilde{x}_a \, d\tilde{x}_a \, d\phi_a \, \tilde{x}_b \, d\tilde{x}_b \, d\phi_b \, dy_a \, dy_b
 f_a\, f_b \>\nonumber \\
& &\times \, \delta\left[M^2-2M^2 \tilde{x}_a \tilde{x}_b 
\cosh (y_a-y_b)+2 M^2 \tilde{x}_a \tilde{x}_b \cos \phi_b\right]
\ea
where $\tilde{x}_a=p_T^a/M$, $\tilde{x}_b=p_T^b/M$ and $y_a$ and $y_b$ 
are the rapidities.
The integrations over the azimuthal angles yield
\ba
\frac{dR}{dM^2}=\frac{M^4\sigma(M^2)}{(2\pi)^5} \int
\tilde{x}_a \, d\tilde{x}_a \, \tilde{x}_b \, d\tilde{x}_b \, dy_a \, dy_b\, 
 f_a \,f_b \,\nonumber  \\ \times\, \left [ 4 \tilde{x}_a^2 \tilde{x}_b^2 -
\left\{2 \tilde{x}_a \tilde{x}_b \cosh (y_a-y_b)-1\right\}^2
\right ]^{-1/2}\,,
\label{gen}
\ea
such that
\begin{eqnarray}
 -1 &  \le & \, \frac{2\tilde{x}_a \tilde{x}_b \, \cosh (y_a-y_b) -1}{2 
\tilde{x}_a \tilde{x}_b} \, \le \, 1
\> , \nonumber\\
 0 & \le & \, \tilde{x}_{a,b}\, \le \,  \infty \> , \\
-\infty & \le & \, y_{a,b} \, \le \, \infty \> . \nonumber
\end{eqnarray}
When $f_a$ and $f_b$ are given by a thermal distribution
\begin{equation}
f_{\rm th}({\bf{p}})=\exp(-E/T)=\exp(-p_T \cosh y/T)\,,
\label{fther}
\end{equation}
the above integral can be performed to obtain  Eq. (\ref{dratequarks}). 

\begin{figure}[tb]
  \begin{center}
  \epsfig{file=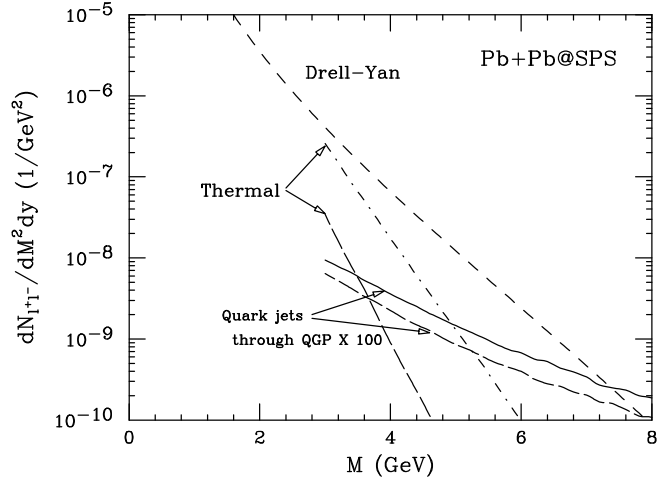,width=8.6cm} 
  \caption{Dilepton spectrum\protect\cite{SGF} for Pb+Pb at $\sqrt{s_{\rm NN}}
  =17.4$ GeV at SPS.  The solid line is the result with $\tau_0$ = 0.2 fm/$c$. 
  The long dashed curves give the 
results when the formation time $\tau_0$ is raised to 0.50 fm/$c$,
thus lowering the temperature. See the quoted reference for details.}
  \label{fig:dilsps}
  \end{center}
\end{figure}

\begin{figure}[tb]
  \begin{center}  
  \epsfig{file=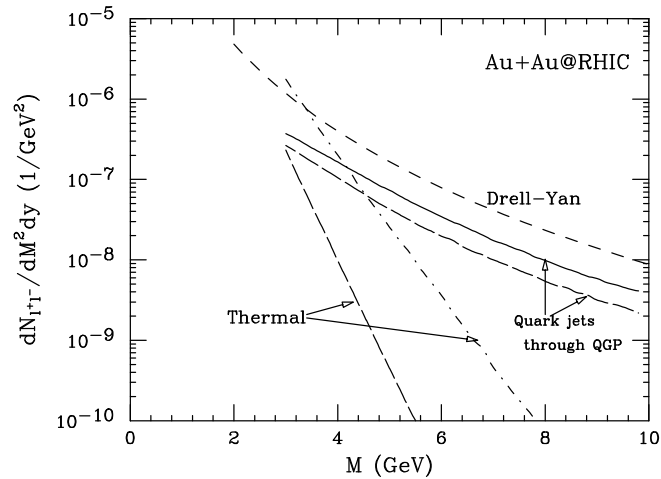,width=8.6cm} 
  \caption{Same as Fig.~\ref{fig:dilsps} for central Au+Au at 
  $\sqrt{s_{NN}}=200$ GeV at RHIC.}
  \label{fig:dilrhic}
  \end{center}
\end{figure}

\begin{figure}[tb]
  \begin{center}
  \epsfig{file=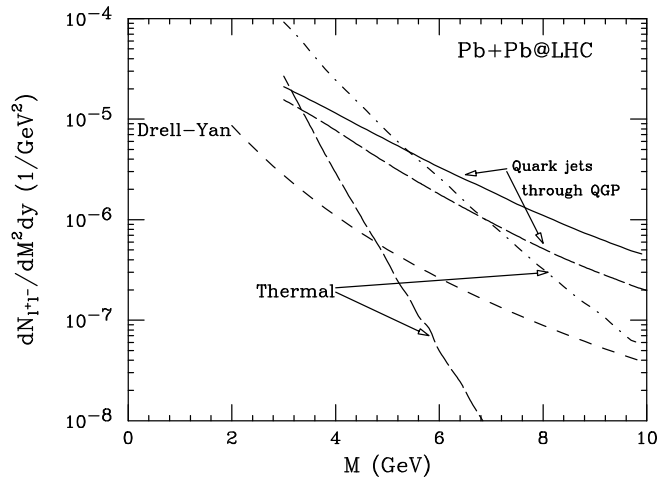,width=8.6cm} 
  \caption{Same as Fig.~\ref{fig:dilsps} for central Pb+Pb at 
  $\sqrt{s_{NN}}=5.5$ TeV at LHC.}
  \label{fig:dillhc}
  \end{center}
\end{figure}

Again assuming a Bjorken-scenario isentropic plasma evolution, one can plot
results for thermal dileptons, dileptons from the Drell-Yan process, and the
dileptons from the passage of quark jets through the plasma for SPS, RHIC, and
LHC respectively. Those constitute figures~\ref{fig:dilsps}, \ref{fig:dilrhic} 
and \ref{fig:dillhc}. 

At SPS energies, we recover (Fig.~\ref{fig:dilsps})
the well known result that the high mass 
dileptons have their origin predominantly in the Drell-Yan process. 
Increasing the formation time from 0.20 fm/$c$ to 
0.50 fm/$c$ --- and thus lowering the
initial temperature by 100 MeV --- drastically alters the thermal 
production (from the dash-dotted curve to the long-dashed one) while
the yield from the proposed jet-plasma interaction, even though
essentially negligible, is reduced by a factor of $\approx$ 2 (from the solid
line to the long-dashed one).

The jet-plasma interaction starts playing an interesting role at
RHIC energies (Fig.~\ref{fig:dilrhic}), as now the corresponding
yield is about only one third of the Drell-Yan contribution, and 
is much larger than the thermal contribution. Again
lowering the initial temperature (now by about 150 MeV) by increasing
the formation time to 0.50 fm/$c$ further enhances the importance 
of the yield due to jet-plasma interaction. This production
is of the same order as that attributed to secondary-secondary 
quark-antiquark annihilation in a dilepton production calculation done 
using an earlier version of
the parton cascade model\cite{gk:93}.

The much higher initial temperatures likely to be attained at the
LHC and the much larger (mini)jet production lead 
to an excess of high mass dileptons from jet-plasma interactions
which can be an order of magnitude greater (at $M=10$ GeV) than that due
to the Drell-Yan process. Again, reducing the initial temperature
by raising the formation time to 0.50 fm/$c$ reduces the jet-plasma
yield by about a factor of 2 while the thermal yield is reduced
far more. Recall that at LHC energies several calculations\cite{kms:92,SriMusMul:97,ramona} 
have reported a thermal yield larger than the Drell-Yan production.
It is found that the jet-plasma interaction enhances the high mass 
dilepton production considerably. The calculations outlined above can be
repeated for a plasma that is not in chemical equilibrium\cite{SGF}. In this
case, conclusions similar to the ones reached above are obtained. Summarising,
it appears that a unique source of high mass dileptons is generated by the
passage of quark jets through the quark-gluon plasma. The contribution is seen
to be the largest at LHC energies, moderate at RHIC energies, and negligible at
SPS energies. The measurement of this radiation could then be added to the list
of QGP signatures, as well as providing a proof of existence for the conditions
suitable for jet-quenching to occur.

Finally, even though this is not a ``direct'' plasma signal,  it is worth mentioning 
that electromagnetic radiation can also serve as a versatile jet-tag. This is
especially useful in environments where the jet is expected to loose energy, or
to be quenched out of existence. This statement holds true for real
photons\cite{wang_pho}, as well as for lepton pairs\cite{SGA}. 

\subsection{Squeezing Lepton Pairs out of Broken Symmetries}
We have seen in the text above that the electromagnetic radiation
measured in nuclear collisions is a precious measure of the in-medium 
photon self-energy. However, in a bath of finite temperature and density, 
new possibilities can manifest themselves. In some sense, the medium
allows for the existence of correlators that vanish identically in the
vacuum. This fact opens mixing channels that were previously closed.
While several studies have sought to investigate the in-medium
properties of hadrons, their mixing with other hadrons has up to now
received little attention. An exception is the case of $\rho - \omega$
\cite{mix1}. This specific mixing may be omitted when dealing with
isospin symmetric nuclear matter, as is done here. Also, one will
concentrate on vector mesons, as they enjoy a privileged relationship
with electromagnetic signals.  First, one describes an exploratory
calculation designed to highlight an eventual signal. The possibility of
$\rho - a_0$ mixing is explored, via nucleon-nucleon excitations in
strongly interacting systems. It is shown that this mixing opens up a new
channel for the dilepton production and may thus induce an additional
peak in the $\phi$ region. A similar mixing exists with the $\sigma -
\omega$\cite{chin,sttw}, but its electromagnetic signatures are much 
smaller\cite{sigom,tdg2}.

For the purposes described above, the interaction Lagrangian can be
written as
\begin{eqnarray}
{\cal L} & = & g_\sigma \bar{\Psi} \phi_\sigma \Psi + g_{a_0} \bar{\Psi}
\phi_{a_0, b} \tau^b \Psi + g_{\omega N N} \bar{\Psi} \gamma_\mu \Psi
\omega^\mu
\nonumber\\
& + & g_\rho \left[ \bar{\Psi} \gamma_\mu \tau^\alpha \Psi +
\frac{\kappa_\rho}{2 m_N} \bar{\Psi} \sigma_{\mu \nu} \tau^\alpha
\Psi
\partial^\nu \right] \rho_\alpha^\mu\ ,
\end{eqnarray}
where $\Psi$, $\phi_\sigma$, $\phi_{a_0}$, $\rho$, and $\omega$
correspond to nucleon, $\sigma$, $a_0$ $\rho$, and $\omega$ fields, and
$\tau_b$ is a Pauli matrix. The values for the coupling parameters are
from [\refcite{bonn}]. The existence of a preferred rest frame essentially 
creates a new vacuum state with quantum numbers different from those of the true
vacuum. An immediate consequence of this fact is that one can now define a mixed
correlator involving scalar and vector current operators: 
$\langle j^{\rm S} j^{\rm V}_\mu \rangle$. This mixed correlator is identically
zero in the true vacuum.
The polarisation vector through which the $a_0$
couples to the $\rho$ via the $N - N$ loop is given by
\begin{eqnarray}
\Pi_\mu (q_0, |\vec{q}\,|) = 2 i g_{a_0} g_\rho \int \frac{d^4 k}{(2 \pi
)^4}\, {\rm Tr} \left[ G(k) \Gamma_\mu G(k+q)\right]\ ,
\label{polariz}
\end{eqnarray}
where 2 is an isospin factor and the vertex for $\rho\/-\/N\/-\/N$ coupling
is
\begin{eqnarray}
\Gamma_\mu = \gamma_\mu - \frac{\kappa_\rho}{2 m_N} \sigma_{\mu \nu}
q^\nu\ .
\end{eqnarray}
In the above $G(k)$ is the in-medium nucleon propagator\cite{sewa}.
For the sake of simplicity, one first uses the density-dependent and
temperature-independent propagator. This approximation will be relaxed
later. With the evaluation of the trace and after a little algebra,
Eq.~\ref{polariz} can be cast in a suggestive form:
\begin{eqnarray}
\Pi_\mu (q_0, |\vec{q}\,|) = \frac{g_{a_0} g_\rho}{\pi^3} \, 2 q^2
\left(2 m_N^* - \frac{\kappa q^2}{2 m_N}\right) \, \int_0^{k_F} \frac{d^3
k}{E^* (k)}\, \frac{k_\mu - \frac{q_\mu}{q^2} (k \cdot q)}{q^4 - 4 (k
\cdot q)^2}\ .
\end{eqnarray}
This leads immediately to two conclusions. First, $q_\mu \Pi^\mu$ = 0.
Second, only two components of the polarisation vector survive after the
angular integration.  This will guarantee that only the longitudinal
component of the $\rho$ couples to the scalar meson, while the
transverse mode remains unaltered. Further note that current
conservation implies that out of the two nonzero components of
$\Pi_\mu$, only one is independent. The new mixing channel will affect
the properties of the mesons in medium, {\it\/i.e.\/} affect their masses and
spectral function. Those aspects will not be discussed at length here,
but details can be found in the literature\cite{sttw,tdg3}. 

\begin{figure}[tb]
  \begin{center}  
  \epsfig{file=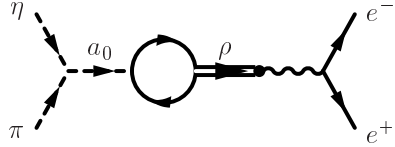,width=5.6cm} 
  \caption{Feynman diagram for the process $\pi + \eta \to e^+ e^-$, which
  proceeds through $\rho-a_0$ mixing.}
  \label{rhoa0Feyn}
  \end{center}
\end{figure}
\begin{figure}
\begin{center}
\epsfig{file=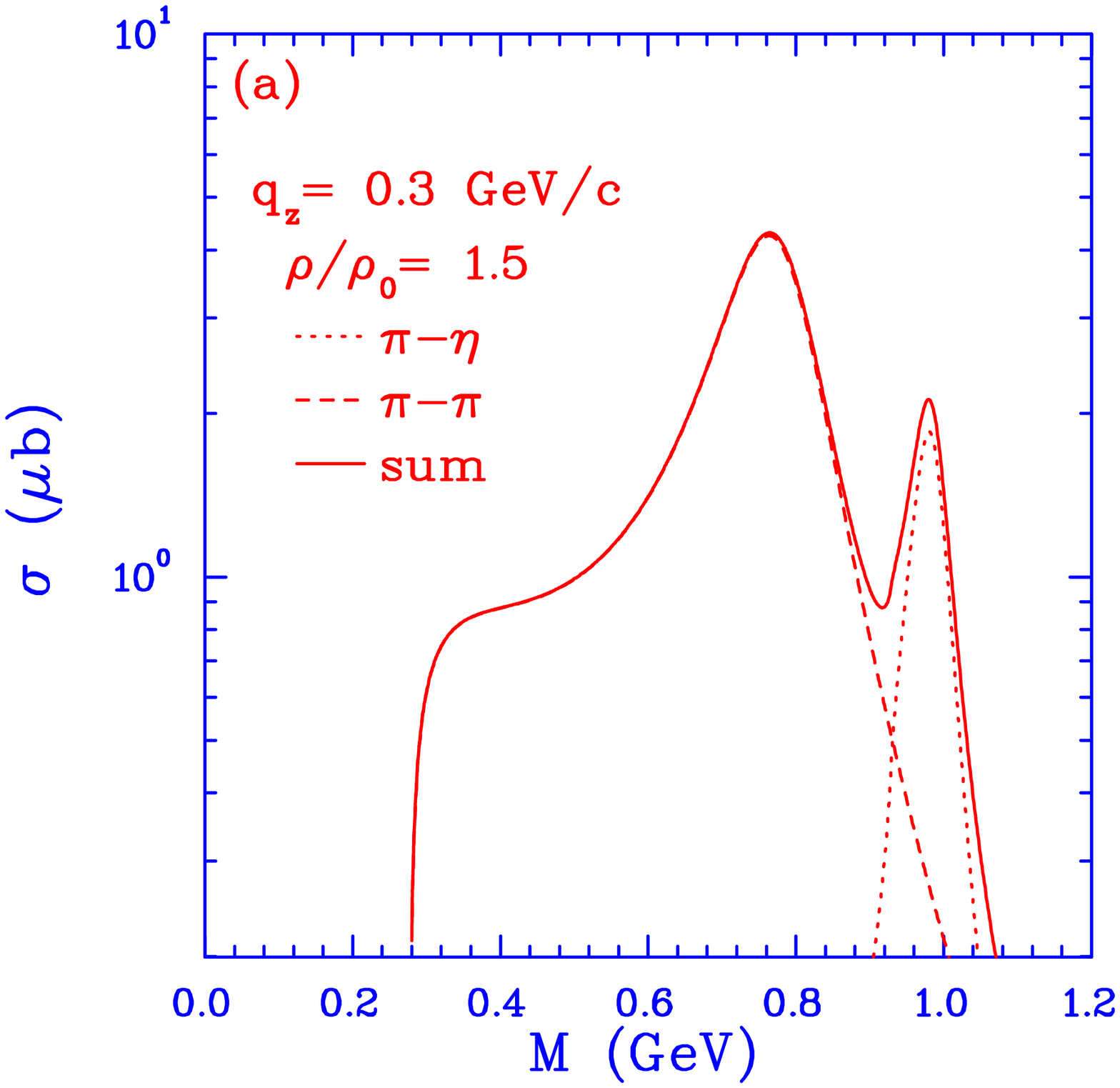,height=5.0cm,angle=0}\epsfig{file=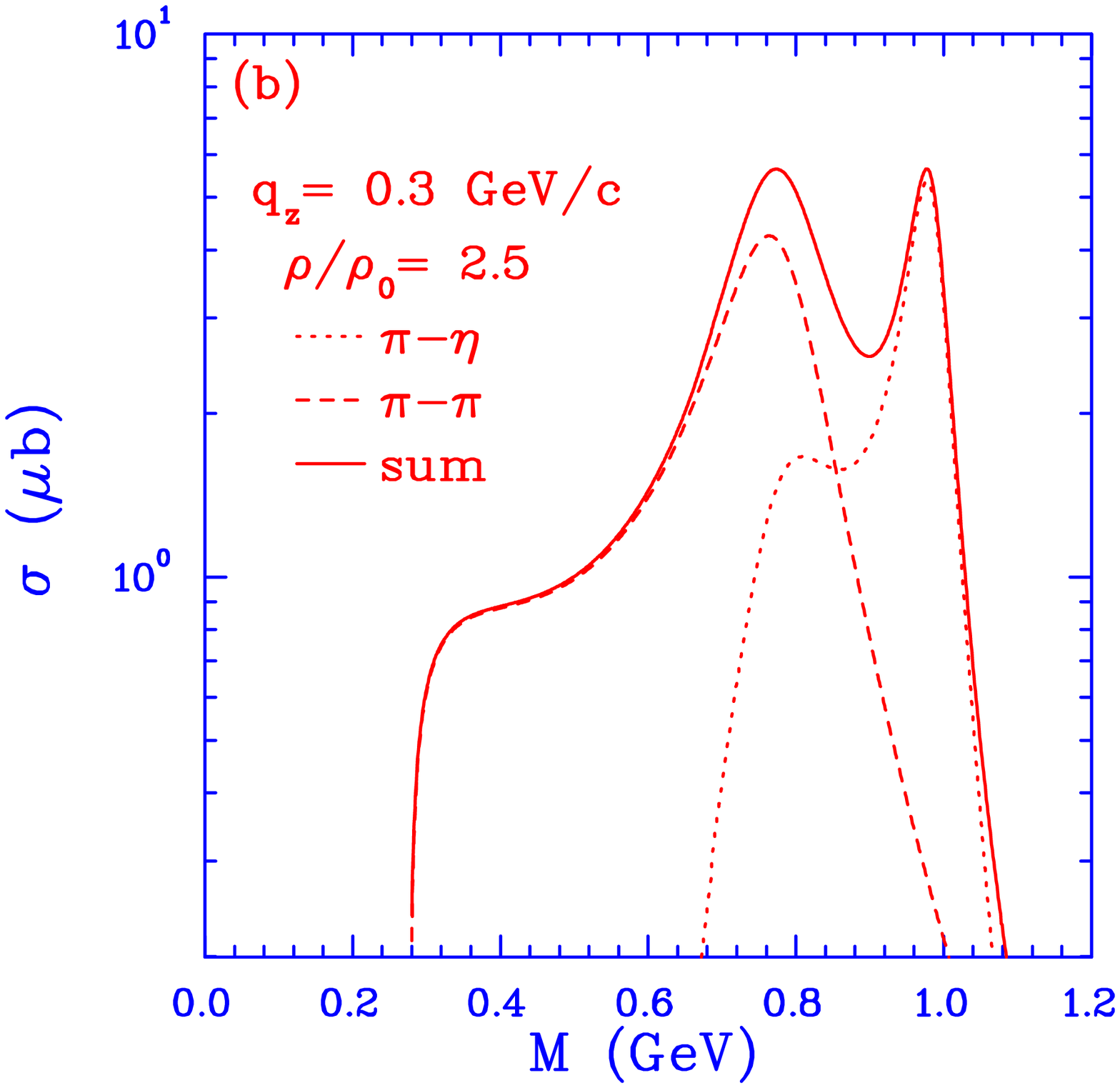,height=5.0cm,angle=0}
\end{center}
 \caption[Dilepton spectrum] 
  {\small Dilepton spectrum induced by $\pi + \pi
\rightarrow e^+ + e^-$ and $\pi + \eta
\rightarrow e^+ + e^-$ considering matter-induced $\rho-a_0$ mixing. (a)  
${\rho}$=1.5${\rho}_0$ (b) ${\rho}$=2.5${\rho}_0$
\label{mixcross}}
\end{figure}

The $\rho-a_0$ mixing opens a new channel in dense nuclear matter, $\pi + \eta
\to e^+ e^-$, which may proceed through {\it\/N-N\/} excitations. The 
Feynman diagram for
this process is shown in Fig.~\ref{rhoa0Feyn}. One may then evaluate cross
sections for the production of lepton pairs. Evaluating the polarisation
in the zero-temperature limit, the cross section for this process is
\ba
\sigma_{\pi \eta \rightarrow e^+ e^-}& =& \frac{4 \pi \alpha^2}{3 q_z^2 M}
\frac {g_{a_0 \pi\eta}^2}{g_\rho^2}
\frac{m_\rho^4}{(M^2 - m_\rho^2)^2 + m_\rho^2 \Gamma^2_\rho(M)}\nonumber \\
& &\times\, 
\frac{1}{(M^2 - m_{a_0})^2 + m_{a_0}^2 \Gamma^2_{a_0}(M)}
\frac{1}{{\sqrt{M^2 - 4 m_\pi^2}}}\,{\mid\Pi_0\mid}^2 \ ,
\ea
where $\Pi_0$ is the zeroth component of the expression in Eq. (\ref{polariz}).
The numerical values for the couplings and the calculation details are to be
found in Ref.~[\refcite{tdg1}]. The cross sections are shown in 
Fig.~\ref{mixcross},
for two different values of the nuclear density. The familiar process $\pi \pi
\to e^+ e^-$ is also plotted to set a scale. 
A noticeable feature of this plot is that the mixing process induces a peak at
$m_{a_0}$ = 0.985 GeV. This constitutes a genuine in-medium effect which is
mostly density-driven: this peak
would be completely absent in vacuum. Furthermore, the contribution to the cross
section at the $a_0$ mass is comparable in magnitude to that of the $\pi \pi$
channel at its peak. Calculations of emission rates where the $T$ = 0
simplification was not made also support this claim\cite{tdg2}. Those mixing
effects will also affect the bulk features of in-medium mesonic
behaviour\cite{tdg3}.  A natural question to ask is whether those
symmetry-breaking effects in the dilepton spectrum should have been observed in
any of the past or present experiments. It turns out that the required baryonic
density is too transient to influence significantly this signal at CERN
energies\cite{teothesis}. However, the HADES\cite{hades} experiment at the 
GSI, in Darmstadt, Germany, has the resolution and the sensitivity to
explore the appropriate invariant mass range. Those heavy ion reactions are performed 
at lower energies, with respect to those of CERN, and thus lead to higher 
baryonic densities which persist longer.   Evidencing
continued theoretical interest at these lower energies,
there are new advancements\cite{elena,mo03} in theory pertinent for the 
dilepton measurements made by the DLS\cite{DLS}, and to be made by HADES. 

There is finally another symmetry-breaking with an electromagnetic
signature that will be mentioned, even though quantitative evaluations are still
in their preliminary stage. At zero temperature, and at finite temperature and 
zero charge density, diagrams in QED that
contain a fermion loop with an odd number of photon vertices  
are cancelled by an equal and opposite contribution coming from the same
diagram with fermion lines running in the opposite direction, this is the 
basic content of Furry's theorem\cite{fur37} (see also [\refcite{itz80,wei95}]). 

In the language of operators we note that these 
diagrams are encountered in the perturbative evaluation of Green's 
functions with an odd  number of gauge field operators:

\[
 \langle 0| A_{\mu1} A_{\mu2} \ldots A_{\mu2n+1}
 |0\rangle . 
\]
 
In QED we know that
 $CA_{\mu}C^{-1} = -A_{\mu} $, where $C$ is the charge conjugation operator. In the case of the
 vacuum $|0\rangle $, we note that $C|0\rangle = |0\rangle$, as the vacuum is 
uncharged. As a result 


\ba
 \langle 0| A_{\mu_1} A_{\mu_2} \ldots A_{\mu_{2n+1}} |0\rangle 
&=&  \langle 0| C^{-1}C A_{\mu_1} C^{-1}C A_{\mu_2} \ldots A_{\mu_{2n+1}} C^{-1} C |0\rangle
\nonumber \\
&=& \langle 0| A_{\mu_1} A_{\mu_2} \ldots A_{\mu_{2n+1}} |0\rangle (-1)^{2n+1}
\nonumber \\
&=& -\langle 0| A_{\mu_1} A_{\mu_2} \ldots A_{\mu_{2n+1}} |0\rangle = 0. \label{fur1}
\ea  

In an equilibrated medium at a temperature $T$, we not only have the expectation of 
the operator on the ground state but on all possible matter 
states weighted by a Boltzmann factor: 

\[
\sum_{n} \langle n| A_{\mu_1} A_{\mu_2} \ldots A_{\mu_{2n+1}} |n\rangle 
e^{-\beta (E_n - \mu Q_n)},
\]
where $\beta = 1/T$ and $\mu$ is a chemical potential.
We are thus calculating the expectation in the grand canonical ensemble.
Here, $C|n\rangle = e^{i\phi}|-n\rangle$, where $|-n\rangle$  is a state in the ensemble with
the same number of antiparticles as there are particles in $|n\rangle$ and vice-versa.
If $\mu = 0$ i.e., the ensemble average displays zero density 
then inserting the operator $C^{-1}C$ as before, we get
\begin{eqnarray}
\langle n| A_{\mu_1} A_{\mu_2} &\ldots& A_{\mu_{2n+1}} |n\rangle 
e^{-\beta E_n} \nonumber \\
&  =& - \langle -n| A_{\mu_1} A_{\mu_2} \ldots A_{\mu_{2n+1}} |-n\rangle 
e^{-\beta E_n}\mbox{}
\end{eqnarray}
The sum over all states will contain the mirror term 
$\langle -n| A_{\mu_1} A_{\mu_2} \ldots A_{\mu_{2n+1}} |{-n}\rangle e^{-\beta E_n} $, with
the same thermal weight, hence

\begin{eqnarray}
\sum_{n} \langle n| A_{\mu_1} A_{\mu_2} \ldots A_{\mu_{2n+1}} |n\rangle 
e^{-\beta E_n } = 0,
\end{eqnarray}
(the expectation over states which are excitations of the vacuum $|0\rangle$ 
will again be zero as in Eq. \ref{fur1}) 
and Furry's theorem still holds. 
However, if
$\mu \neq 0$ 
($\Rightarrow$ unequal number of particles and antiparticles ) 
then


\begin{eqnarray}
\langle n| A_{\mu_1} A_{\mu_2} &\ldots& A_{\mu_{2n+1}} |n\rangle 
e^{-\beta (E_n - \mu Q_n)}\nonumber \\
& =& - \langle -n| A_{\mu_1} A_{\mu_2} \ldots A_{\mu_{2n+1}} |-n\rangle 
e^{-\beta (E_n - \mu Q_n)},
\end{eqnarray}

\noindent the mirror term this time is 
$
 \langle -n| A_{\mu_1} A_{\mu_2} \ldots A_{\mu_{2n+1}} |-n\rangle 
e^{-\beta (E_n + \mu Q_n)},
$
with a different thermal weight, thus 


\begin{eqnarray}
\sum_{n} \langle n| A_{\mu_1} A_{\mu_2} \ldots A_{\mu_{2n+1}} |n\rangle 
e^{-\beta (E_n - \mu Q_n)} \neq 0.
\end{eqnarray}

The sum over all medium states which leads to the expectation value is no longer zero. This 
represents the breaking of Furry's theorem by the medium. Note that this occurs only in 
media with non-zero density or chemical potential. Summarising, if the medium
contains a net charge, such that it breaks $C$ spontaneously, Green's functions
that vanish identically in the vacuum (or in a neutral medium) can survive.
Making the simplest possible extension to QCD, one may replace two of the
photons with gluons. This enables processes like $g g \to \ell^+ \ell^-$, where
the gluon fusion proceeds through a quark (antiquark) loop. This channel is
exciting for the following reason: it offers a direct electromagnetic signature
of early gluon populations. This represents pristine information on the state of
the many-body system. Calculations are technically involved\cite{majbou}, but
results are finally forthcoming\cite{bmg03}.

\section{Conclusions}

This ends our survey of the use of electromagnetic signals as probes of strongly-interacting 
relativistic many-body dynamics.   The supporting framework 
has been relativistic quantum field theory, generalised to finite
temperatures and densities, in order to formulate very general computational
tools for estimating production rates from heated and compressed
nuclear systems.  The focus was on 
establishing within equilibrium circumstances, the number of
radiated electromagnetic quanta per unit volume per unit time.  For then, one
can, and did, take the rate and evolve according to some ``best guess''
scenarios for the expansion dynamics in heavy ion collisions realized
at facilities at SPS and RHIC.

Dilepton experimental circumstances resembling those expected at the 
CERN SPS have been modeled in a variety of ways.  Comparisons of theory 
and experiment 
have been fruitful in terms of suggesting a consistent picture of modified
vector meson spectral properties.  These modification are truly
collective nuclear effects.  Tremendous advancement in our 
understanding of the way in which nuclear matter responds when it
is forced near the phase boundary between hadrons and quarks has come from 
these pursuits.  To
mention some specific achievements, one notes that the rho spectral
function has been essentially measured at finite energy density and
has been shown to be significantly modified from its vacuum
structure.  This is a significant achievement. 

The photon studies have been exceptionally fruitful too.  Theory
has advanced from a stage where rate estimates where plagued with
infrared singularities, to first establishing regulated lowest-order
results with new computational techniques and tools.  The hard-thermal-loop
approach is useful not only for photon production, but other
studies in hot gauge theories and for a variety of observables.
Next, one witnessed an impressive effort to understand 
the photon self-energy up to the many-loop order and
including multiple-scattering effects within the medium.  The
celebrated result is a complete lowest order photon production rate
from finite temperature QCD that is stable and reliable.  That
too is a nontrivial accomplishment.

When the QCD rates and hadronic rates are used to predict photon 
yields and then compared with experiments WA80 and WA98 from CERN, there is 
consistency, if not discriminatory features.  One can say that the results
strongly suggest thermal emission from a fireball at roughly 200 MeV 
temperature.   Have we observed the QGP? The investigations reported on in this
work only contain hints of an answer. As mentioned previously, the radiation
from the partonic phase is present in the analyses, but does not constitute a
large portion of the overall signal. Fortunately, this state of affairs is
directly related to probed temperatures and to the space-time volume occupied by
the plasma. Both those are expected to increase in the current and future
generations of collider experiments. The only indirect proof is that, in many
dynamical simulations, it is unavoidable for the initial phase to be elsewhere
in the phase diagram than in the deconfined region. This is a consistency
requirement brought about by our current knowledge of the equation of state.
This however will change and evolve, especially with the experience gained with
non-perturbative approaches which will in turn guarantee a better focused picture of the
quasiparticle nature of the partonic sector. Have we observed chiral symmetry
restoration? The approach to chiral symmetry is closely related to the
properties of the in-medium spectral densities\cite{kap_shu}. As the constraints of
the Weinberg sum rules are extended to finite temperatures, they require a
degeneracy of the vector and axial-vector correlators in the symmetric limit.
This demand, however, can be satisfied in several ways\cite{kap_shu}. Thus, a
verdict on the
status of chiral symmetry restoration is being hindered by the difficulty to
access the axial-vector correlator unambiguously. More theoretical work needs to
be done in that respect as well, in order to provide a unified calculation with
a credible degree of sophistication. 
It should finally be mentioned that RHIC and the LHC will have the intensity to
make possible Hanbury-Brown-Twiss interferometry measurements of direct photons.
The theory of this observable is well-developed\cite{HBT}, and measurements of
the correlation functions are expected to place constraints on the space-time
extent of the photon-emitting sources\cite{HBT2}. 

If the estimates brought about by dynamical approaches and by analyses of hadrons abundances
are reliable, we have just grazed the phase boundary of the deconfined sector at
the SPS. This assertion is not in conflict with the evidence obtained from
measurements of electromagnetic observables. Then, RHIC, and the LHC,
should soon signal bold incursions into a new territory.  

\section*{Acknowledgements}
It is a pleasure to acknowledge helpful comments from, and discussions with, 
P. Aurenche, J.-e. Alam, A. Bourque, E. Bratkovskaya, A. Dutt-Mazumder, S. Gao,
F. G\'elis, 
P. Jaikumar, B. K\"ampfer, J. Kapusta, V. Koch, I. Kvasnikova, 
G. D.  Moore, R. Rapp, 
D. Srivastava, O. Teodorescu, Y. Tserruya, and S. Turbide. 
C. G. thanks J. Bruce, J. Entwistle, J. Pastorius, Ch. Squire, and G. 
Willis for inspiration.
This work has been supported in part by the Natural Sciences
and Engineering Research Council of Canada, in part by 
the the Fonds Nature et Technologies of Quebec, and in
part by the National Science Foundation under grant 
number PHY-0098760.

\end{document}